\documentclass[twocolumn]{aastex61}

\usepackage[utf8]{inputenc}
\usepackage[titletoc,title]{appendix}
\usepackage{CJKutf8}
\usepackage{float}
\newcommand{\be}{\begin{equation}}
\newcommand{\ee}{\end{equation}}
\newcommand{\thisMainObj}{\mbox{2015 BP$_{519}$}}

\newcommand{\listOfTnos}{2003 VB$_{12}$ (Sedna), 
 2004 VN$_{112}$,
2007 TG$_{422}$,
2010 GB$_{174}$,
2012 VP$_{113}$,
2013 FT$_{28}$,
2013 RF$_{98}$,
2013 SY$_{99}$,
2014 FE$_{72}$,
2014 SR$_{349}$,
2015 GT$_{50}$,
2015 KG$_{163}$, and
2015 RX$_{245}$}
\newcommand{\listOfTnosWithCitations}{2003 VB$_{12}$ \citep[known as Sedna,][]{sedna}, 
2004 VN$_{112}$ (MPC),
2007 TG$_{422}$ (MPC),
2010 GB$_{174}$ \citep{gb1074},
2012 VP$_{113}$ \citep{st14},
2013 FT$_{28}$ \citep{ST_tnos},
2013 RF$_{98}$ (MPC),
2013 SY$_{99}$ \citep{2017AJ....153..262B},
2014 FE$_{72}$ \citep{ST_tnos},
2014 SR$_{349}$ \citep{ST_tnos},
2015 GT$_{50}$ \citep{2017AJ....154...50S},
2015 KG$_{163}$ \citep{2017AJ....154...50S}, and
2015 RX$_{245}$ \citep{2017AJ....154...50S}}

\newcommand{\thisMainObjSMA}{450 AU}
\newcommand{\thisMainObjECC}{0.92}
\newcommand{\thisMainObjINC}{54 degrees}

\usepackage{xcolor}
\usepackage[titletoc,title]{appendix}

\usepackage{amsmath}
\usepackage{graphicx}
\newcommand{\quantityE}{\sum^{N}_{j}(m_{j}a_{j}^{4})}
\newcommand{\quantityC}{\sum^{N}_{j}(m_{j}a_{j}^{2})}

\usepackage{verbatim}
\usepackage{booktabs}
\begin{document}
\shorttitle{Discovery and Dynamical Analysis of an Extreme TNO }
\shortauthors{Becker et al. }

\title{Discovery and Dynamical Analysis of an Extreme Trans-Neptunian Object
with a High Orbital Inclination}

\correspondingauthor{Juliette Becker}
\email{jcbecker@umich.edu}

\author[0000-0002-7733-4522]{J.~C.~Becker}
\altaffiliation{NSF Graduate Research Fellow}
\affiliation{Department of Astronomy, University of Michigan, Ann Arbor, MI 48109, USA}
\author[0000-0001-7721-6457]{T.~Khain}
\affiliation{Department of Physics, University of Michigan, Ann Arbor, MI 48109, USA}
\author[0000-0002-6126-8487]{S.~J.~Hamilton}
\altaffiliation{NSF Graduate Research Fellow}
\affiliation{Department of Physics, University of Michigan, Ann Arbor, MI 48109, USA}
\author[0000-0002-8167-1767]{F.~C.~Adams}
\affiliation{Department of Physics, University of Michigan, Ann Arbor, MI 48109, USA}
\affiliation{Department of Astronomy, University of Michigan, Ann Arbor, MI 48109, USA}
\author[0000-0001-6942-2736]{D.~W.~Gerdes}
\affiliation{Department of Physics, University of Michigan, Ann Arbor, MI 48109, USA}
\affiliation{Department of Astronomy, University of Michigan, Ann Arbor, MI 48109, USA}
\author[0000-0002-8906-2835]{L.~Zullo}
\affiliation{Department of Physics, University of Michigan, Ann Arbor, MI 48109, USA}
\author[0000-0002-8906-2835]{K.~Franson}
\affiliation{Department of Physics, University of Michigan, Ann Arbor, MI 48109, USA}
\author[0000-0003-3130-2282]{S.~Millholland}
\affiliation{Department of Astronomy, Yale University, New Haven, CT 06511, USA}
\author{G.~M.~Bernstein}
\affiliation{Department of Physics and Astronomy, University of Pennsylvania, Philadelphia, PA 19104, USA}
\author[0000-0003-2764-7093]{M.~Sako}
\affiliation{Department of Physics and Astronomy, University of Pennsylvania, Philadelphia, PA 19104, USA}
\author[0000-0003-0743-9422]{P.~Bernardinelli}
\affiliation{Department of Physics and Astronomy, University of Pennsylvania, Philadelphia, PA 19104, USA}
\author{K. Napier}
\affiliation{Department of Physics, University of Michigan, Ann Arbor, MI 48109, USA}
\affiliation{Siena College, Loudonville, NY 12211, USA}
\author[0000-0002-2486-1118]{L. Markwardt} 
\altaffiliation{NSF Graduate Research Fellow}
\affiliation{Department of Astronomy, University of Michigan, Ann Arbor, MI 48109, USA}
\author[0000-0001-7737-6784]{Hsing~Wen~Lin (\begin{CJK*}{UTF8}{gbsn}
林省文\end{CJK*})}
\affiliation{Department of Physics, University of Michigan, Ann Arbor, MI 48109, USA}
\author[0000-0003-0072-6736]{W.~Wester}
\affiliation{Fermi National Accelerator Laboratory, P. O. Box 500, Batavia, IL 60510, USA}

\author{F.~B.~Abdalla}
\affiliation{Department of Physics \& Astronomy, University College London, Gower Street, London, WC1E 6BT, UK}
\affiliation{Department of Physics and Electronics, Rhodes University, PO Box 94, Grahamstown, 6140, South Africa}
\author{S.~Allam}
\affiliation{Fermi National Accelerator Laboratory, P. O. Box 500, Batavia, IL 60510, USA}
\author{J.~Annis}
\affiliation{Fermi National Accelerator Laboratory, P. O. Box 500, Batavia, IL 60510, USA}
\author{S.~Avila}
\affiliation{Institute of Cosmology \& Gravitation, University of Portsmouth, Portsmouth, PO1 3FX, UK}
\author{E.~Bertin}
\affiliation{CNRS, UMR 7095, Institut d'Astrophysique de Paris, F-75014, Paris, France}
\affiliation{Sorbonne Universit\'es, UPMC Univ Paris 06, UMR 7095, Institut d'Astrophysique de Paris, F-75014, Paris, France}
\author{D.~Brooks}
\affiliation{Department of Physics \& Astronomy, University College London, Gower Street, London, WC1E 6BT, UK}
\author{A.~Carnero~Rosell}
\affiliation{Laborat\'orio Interinstitucional de e-Astronomia - LIneA, Rua Gal. Jos\'e Cristino 77, Rio de Janeiro, RJ - 20921-400, Brazil}
\affiliation{Observat\'orio Nacional, Rua Gal. Jos\'e Cristino 77, Rio de Janeiro, RJ - 20921-400, Brazil}
\author{M.~Carrasco~Kind}
\affiliation{Department of Astronomy, University of Illinois at Urbana-Champaign, 1002 W. Green Street, Urbana, IL 61801, USA}
\affiliation{National Center for Supercomputing Applications, 1205 West Clark St., Urbana, IL 61801, USA}
\author{J.~Carretero}
\affiliation{Institut de F\'{\i}sica d'Altes Energies (IFAE), The Barcelona Institute of Science and Technology, Campus UAB, 08193 Bellaterra (Barcelona) Spain}
\author{C.~E.~Cunha}
\affiliation{Kavli Institute for Particle Astrophysics \& Cosmology, P. O. Box 2450, Stanford University, Stanford, CA 94305, USA}
\author{C.~B.~D'Andrea}
\affiliation{Department of Physics and Astronomy, University of Pennsylvania, Philadelphia, PA 19104, USA}
\author{L.~N.~da Costa}
\affiliation{Laborat\'orio Interinstitucional de e-Astronomia - LIneA, Rua Gal. Jos\'e Cristino 77, Rio de Janeiro, RJ - 20921-400, Brazil}
\affiliation{Observat\'orio Nacional, Rua Gal. Jos\'e Cristino 77, Rio de Janeiro, RJ - 20921-400, Brazil}
\author{C.~Davis}
\affiliation{Kavli Institute for Particle Astrophysics \& Cosmology, P. O. Box 2450, Stanford University, Stanford, CA 94305, USA}
\author{J.~De~Vicente}
\affiliation{Centro de Investigaciones Energ\'eticas, Medioambientales y Tecnol\'ogicas (CIEMAT), Madrid, Spain}
\author{H.~T.~Diehl}
\affiliation{Fermi National Accelerator Laboratory, P. O. Box 500, Batavia, IL 60510, USA}
\author{P.~Doel}
\affiliation{Department of Physics \& Astronomy, University College London, Gower Street, London, WC1E 6BT, UK}
\author{T.~F.~Eifler}
\affiliation{Department of Astronomy/Steward Observatory, 933 North Cherry Avenue, Tucson, AZ 85721-0065, USA}
\affiliation{Jet Propulsion Laboratory, California Institute of Technology, 4800 Oak Grove Dr., Pasadena, CA 91109, USA}
\author{B.~Flaugher}
\affiliation{Fermi National Accelerator Laboratory, P. O. Box 500, Batavia, IL 60510, USA}
\author{P.~Fosalba}
\affiliation{Institut d'Estudis Espacials de Catalunya (IEEC), 08193 Barcelona, Spain}
\affiliation{Institute of Space Sciences (ICE, CSIC),  Campus UAB, Carrer de Can Magrans, s/n,  08193 Barcelona, Spain}
\author{J.~Frieman}
\affiliation{Fermi National Accelerator Laboratory, P. O. Box 500, Batavia, IL 60510, USA}
\affiliation{Kavli Institute for Cosmological Physics, University of Chicago, Chicago, IL 60637, USA}
\author{J.~Garc\'ia-Bellido}
\affiliation{Instituto de Fisica Teorica UAM/CSIC, Universidad Autonoma de Madrid, 28049 Madrid, Spain}
\author{E.~Gaztanaga}
\affiliation{Institut d'Estudis Espacials de Catalunya (IEEC), 08193 Barcelona, Spain}
\affiliation{Institute of Space Sciences (ICE, CSIC),  Campus UAB, Carrer de Can Magrans, s/n,  08193 Barcelona, Spain}
\author{D.~Gruen}
\affiliation{Kavli Institute for Particle Astrophysics \& Cosmology, P. O. Box 2450, Stanford University, Stanford, CA 94305, USA}
\affiliation{SLAC National Accelerator Laboratory, Menlo Park, CA 94025, USA}
\author{R.~A.~Gruendl}
\affiliation{Department of Astronomy, University of Illinois at Urbana-Champaign, 1002 W. Green Street, Urbana, IL 61801, USA}
\affiliation{National Center for Supercomputing Applications, 1205 West Clark St., Urbana, IL 61801, USA}
\author[0000-0003-3023-8362]{J.~Gschwend}
\affiliation{Laborat\'orio Interinstitucional de e-Astronomia - LIneA, Rua Gal. Jos\'e Cristino 77, Rio de Janeiro, RJ - 20921-400, Brazil}
\affiliation{Observat\'orio Nacional, Rua Gal. Jos\'e Cristino 77, Rio de Janeiro, RJ - 20921-400, Brazil}
\author{G.~Gutierrez}
\affiliation{Fermi National Accelerator Laboratory, P. O. Box 500, Batavia, IL 60510, USA}
\author{W.~G.~Hartley}
\affiliation{Department of Physics \& Astronomy, University College London, Gower Street, London, WC1E 6BT, UK}
\affiliation{Department of Physics, ETH Zurich, Wolfgang-Pauli-Strasse 16, CH-8093 Zurich, Switzerland}
\author{D.~L.~Hollowood}
\affiliation{Santa Cruz Institute for Particle Physics, Santa Cruz, CA 95064, USA}
\author{K.~Honscheid}
\affiliation{Center for Cosmology and Astro-Particle Physics, The Ohio State University, Columbus, OH 43210, USA}
\affiliation{Department of Physics, The Ohio State University, Columbus, OH 43210, USA}
\author{D.~J.~James}
\affiliation{Harvard-Smithsonian Center for Astrophysics, Cambridge, MA 02138, USA}
\author{K.~Kuehn}
\affiliation{Australian Astronomical Observatory, North Ryde, NSW 2113, Australia}
\author{N.~Kuropatkin}
\affiliation{Fermi National Accelerator Laboratory, P. O. Box 500, Batavia, IL 60510, USA}
\author[0000-0001-9856-9307]{M.~A.~G. ~Maia}
\affiliation{Laborat\'orio Interinstitucional de e-Astronomia - LIneA, Rua Gal. Jos\'e Cristino 77, Rio de Janeiro, RJ - 20921-400, Brazil}
\affiliation{Observat\'orio Nacional, Rua Gal. Jos\'e Cristino 77, Rio de Janeiro, RJ - 20921-400, Brazil}
\author{M.~March}
\affiliation{Department of Physics and Astronomy, University of Pennsylvania, Philadelphia, PA 19104, USA}
\author{J.~L.~Marshall}
\affiliation{George P. and Cynthia Woods Mitchell Institute for Fundamental Physics and Astronomy, and Department of Physics and Astronomy, Texas A\&M University, College Station, TX 77843,  USA}
\author{F.~Menanteau}
\affiliation{Department of Astronomy, University of Illinois at Urbana-Champaign, 1002 W. Green Street, Urbana, IL 61801, USA}
\affiliation{National Center for Supercomputing Applications, 1205 West Clark St., Urbana, IL 61801, USA}
\author{R.~Miquel}
\affiliation{Instituci\'o Catalana de Recerca i Estudis Avan\c{c}ats, E-08010 Barcelona, Spain}
\affiliation{Institut de F\'{\i}sica d'Altes Energies (IFAE), The Barcelona Institute of Science and Technology, Campus UAB, 08193 Bellaterra (Barcelona) Spain}
\author[0000-0003-2120-1154]{R.~L.~C. Ogando}
\affiliation{Laborat\'orio Interinstitucional de e-Astronomia - LIneA, Rua Gal. Jos\'e Cristino 77, Rio de Janeiro, RJ - 20921-400, Brazil}
\affiliation{Observat\'orio Nacional, Rua Gal. Jos\'e Cristino 77, Rio de Janeiro, RJ - 20921-400, Brazil}
\author{A.~A.~Plazas}
\affiliation{Jet Propulsion Laboratory, California Institute of Technology, 4800 Oak Grove Dr., Pasadena, CA 91109, USA}
\author{E.~Sanchez}
\affiliation{Centro de Investigaciones Energ\'eticas, Medioambientales y Tecnol\'ogicas (CIEMAT), Madrid, Spain}
\author{V.~Scarpine}
\affiliation{Fermi National Accelerator Laboratory, P. O. Box 500, Batavia, IL 60510, USA}
\author{R.~Schindler}
\affiliation{SLAC National Accelerator Laboratory, Menlo Park, CA 94025, USA}
\author{I.~Sevilla-Noarbe}
\affiliation{Centro de Investigaciones Energ\'eticas, Medioambientales y Tecnol\'ogicas (CIEMAT), Madrid, Spain}
\author{M.~Smith}
\affiliation{School of Physics and Astronomy, University of Southampton,  Southampton, SO17 1BJ, UK}
\author{R.~C.~Smith}
\affiliation{Cerro Tololo Inter-American Observatory, National Optical Astronomy Observatory, Casilla 603, La Serena, Chile}
\author{M.~Soares-Santos}
\affiliation{Brandeis University, Physics Department, 415 South Street, Waltham MA 02453}
\author{F.~Sobreira}
\affiliation{Instituto de F\'isica Gleb Wataghin, Universidade Estadual de Campinas, 13083-859, Campinas, SP, Brazil}
\affiliation{Laborat\'orio Interinstitucional de e-Astronomia - LIneA, Rua Gal. Jos\'e Cristino 77, Rio de Janeiro, RJ - 20921-400, Brazil}
\author{E.~Suchyta}
\affiliation{Computer Science and Mathematics Division, Oak Ridge National Laboratory, Oak Ridge, TN 37831}
\author{M.~E.~C.~Swanson}
\affiliation{National Center for Supercomputing Applications, 1205 West Clark St., Urbana, IL 61801, USA}
\author{A.~R.~Walker}
\affiliation{Cerro Tololo Inter-American Observatory, National Optical Astronomy Observatory, Casilla 603, La Serena, Chile}


\collaboration{(DES Collaboration)}

    


\begin{abstract}
We report the discovery and dynamical analysis of \thisMainObj, an extreme Trans-Neptunian Object detected by the Dark Energy Survey at a heliocentric distance of 55~AU, perihelion of $\sim$36~AU, and absolute magnitude $H_r=4.3$. The current orbit, determined from an 1110-day observational arc, has semi-major axis $a\approx$ \thisMainObjSMA, eccentricity $e\approx$ \thisMainObjECC, and inclination $i\approx$ \thisMainObjINC. With these orbital elements, \thisMainObj\ is the most extreme TNO discovered to date, as quantified by the reduced Kozai action, $\eta_{0} = (1-e^{2})^{1/2} \cos{i}$, which is a conserved quantity at fixed semi-major axis $a$ for axisymmetric perturbations. We discuss the orbital stability and evolution of this object, and find that under the influence of the four known giant planets \thisMainObj\ displays rich dynamical behavior, including rapid diffusion in semi-major axis and more constrained variations in eccentricity and inclination. We also consider the long term orbital stability and evolutionary behavior within the context of the Planet Nine hypothesis, and find that \thisMainObj\ adds to the circumstantial evidence for the existence of this proposed new member of the solar system, as it would represent the first member of the population of high-$i$, $\varpi$-shepherded TNOs.
\end{abstract}


\section{Introduction}
\label{sec:intro}

The most extreme members of any dynamical class of solar system objects serve as particularly acute test cases for theories of our solar system's formation and evolution. In particular, Trans-Neptunian Objects (TNOs) with very large semi-major axes probe the most distant observable
regions of the solar system, aiding to reveal the migration histories of the giant planets. Very high inclination TNOs and centaurs mostly remain puzzling. 
Both classes of objects may also be dynamically influenced by distant, yet-unseen perturbers. Indeed, the apparent clustering in orbital and physical space
of the so-called ``extreme TNOs" with $a>250$~AU and perihelion distances $q>30$~AU was used by \citet{bb16} to argue for the existence of a distant super-Earth known as Planet Nine. 

The 13 currently known extreme TNOs have an average orbital inclination of 17.3$^{\circ}$. The most highly-inclined of these objects, 2013 RF$_{98}$, was discovered in our earlier work \citep{2016MNRAS.460.1270D} and has an inclination of 29.6$^{\circ}$, consistent with other members of the scattered disk population.  
In this work, we report the discovery by the Dark Energy Survey of \thisMainObj, a TNO with a semi-major axis of \thisMainObjSMA\ (the sixth-largest among known TNOs), an eccentricity of \thisMainObjECC, and a remarkable inclination of \thisMainObjINC. The orbital elements of this object make it the ``most extreme" of the extreme TNOs, in a sense that we make precise in Section~\ref{sec:extreme}. With a perihelion distance of $q=35.249 \pm 0.078$~AU, it may also be the first purely trans-Neptunian member of the Planet-Nine-induced high-inclination population first predicted in \citet{2017AJ....154..229B}.

Objects in the outer solar system populate several distinct dynamical categories \citep{2008ssbn.book...43G}. Cold classical Kuiper Belt Objects (CKBOs) are dynamically cool, with perihelion distances greater than 40 AU, low orbital eccentricities, and low orbital inclinations \citep{2000Natur.407..979T,2005AJ....129.1117E}. The orbits of these objects are not controlled by dynamical interactions with Neptune, and they may originate from material left over from the formation of the solar system. 
On the contrary, hot classical KBOs as well as resonant KBOs are believed to have been placed in the trans-Neptunian region from smaller original heliocentric distances. 
Another class of objects have orbits that are perturbed significantly through scattering interactions with Neptune \citep{1997Sci...276.1670D,2002Icar..157..269G}. Yet another set of objects have high eccentricities, but also have sufficiently large perihelia that they are not influenced by either scattering or resonant interactions with Neptune.  

Recently, a new subset of objects has attracted considerable attention. The TNOs with semi-major axes $a>150$ AU and perihelia distances beyond 30 AU  were found in \citet{st14} to exhibit a clustering in their argument of perihelion, $\omega$. \citet{bb16} subsequently noted that this clustering persists in physical space (as measured by the longitude of perihelion $\varpi$, where $\varpi = \omega + \Omega$, where $\Omega$ is the longitude of ascending node). \citet{st14} noted that one explanation for the clustering might be a ninth planet, and \citet{bb16} suggested that the existence of a ninth planet of about 10 Earth masses in the outer solar system could explain the apparent alignment of large-semi-major axis objects \citep{bb16}. The motions of objects with $a>250$ AU would in this case be dominated by Planet Nine, while TNOs falling in the intermediate regime, with $a=150-250$ AU, may experience differing degrees of influence from Planet Nine. The TNOs with $a>250$ AU constitute the ``extreme'' TNOs, or ETNOs.
The evidence and consequences of the Planet Nine hypothesis have been explored in previous literature from both dynamical and observational perspectives \citep{bbincl,2016ApJ...823L...3L, 2016ApJ...824L..22M,2016MNRAS.460L.109M,2016ApJ...825...33K,2016ApJ...826...64B, 2016ApJ...827L..24C,2016AJ....152...80H,2016AJ....152...94H,2016MNRAS.462.1972D,ST_tnos,2017AJ....153...63S, sarah, 2017CeMDA.129..329S,2017AJ....154...61B, 2017AJ....154..229B, 2017arXiv171206547H,2017MNRAS.472L..75P, 2018MNRAS.475.4609E,tali}. 
The Planet Nine hypothesis has been invoked to explain the detachment of perihelia distance for the most distant class of TNOs \citep{tali}, the 6 degree solar obliquity \citep{2016AJ....152..126B,2017AJ....153...27G}, and the existence of highly inclined objects in the outer solar system \citep{bbincl}.
The subset of objects discovered so far to have semi-major axis greater than 250 AU and perihelion distances greater than 30 AU (the extreme TNOs) includes \listOfTnosWithCitations. The orbital elements of these objects are listed in the Appendix for reference.
These objects have inclinations ranging from nearly zero up to a maximum of about 30 degrees. 

The orbital inclinations of these high-$a$ objects are of particular interest dynamically. \citet{2009ApJ...697L..91G} discovered 2008~KV$_{42}$ (Drac), the first retrograde Centaur (where a Centaur is an object with a semi-major axis between 5 AU and 30 AU, placing its orbit in the region of the solar system containing the gas giants). This object does not appear to be primordial and could imply the existence of a reservoir of high inclination TNOs. The discovery of the retrograde centaur 2011~KT$_{19}$ (Niku) \citep{2041-8205-827-2-L24} added to the small collection of such objects, and suggested that they may cluster in a common orbital plane. 

\citet{bb16} predicted that Planet Nine could create such a supply of objects by sourcing them from a more distant population of high-inclination orbits, which are in turn generated by Planet Nine.
\citet{2017AJ....154..229B} presented a dynamical model for the orbital evolution of high-inclination, long period ($a>250$ AU) objects and compared the model to the existing high-$a$, high-$i$ objects. However, the objects then known to reside in that population have perihelia $q<30$ AU, and thus experience orbit crossing with the giant planets, complicating their ability to test the Planet Nine hypothesis. 
To better test this particular prediction of the Planet Nine model, high-$a$, high-$i$ objects with perihelion $q>30$ AU are needed. 

Apart from the Planet Nine debate, the continued discovery of new objects in the outer solar system enables a better understanding of how the solar system arrived at its present state. For this reason, many groups have conducted surveys to increase the census of objects known in the outer solar system and better understand their properties \citep[including][]{2002AJ....123.2083M, 2005AJ....129.1117E, 2010A&A...518L.146M, 2010A&A...518L.147L,2010A&A...518L.148L, 2013A&A...555A..15F,2016AJ....152...70B,2016AJ....152..147L}. Data from other surveys or archival data sets have also been utilized to enable solar system science \citep{2008AJ....136...83F, 2012Icar..218..571S, 2014ApJS..211...17A}. 
The Dark Energy Survey (DES, \citealt{des_survey}) follows in these footsteps, enabling study of new populations including high-inclination objects like \thisMainObj.

This paper is organized as follows. We start with a description of the discovery of \thisMainObj\ by DES in Section \ref{sec:discovery}. Given the extreme status of this object, Section \ref{sec:characterization} considers its dynamical status using a secular approach, starting with an analytic treatment of the problem. The dynamics of this object are rich and complicated, so that a complete characterization requires full N-body numerical simulations to include interactions with Neptune and the other giant planets, as well as other complexities. These numerical simulations are presented in Section \ref{sec:numericalknown} for the dynamics of \thisMainObj\ in the context of the currently known solar system. Section \ref{sec:planetnine} then considers the dynamics of this new object in the presence of the proposed Planet Nine. The paper concludes with a discussion of the implications (Section \ref{sec:discussion}) and a summary of the results (Section \ref{sec:conclude}).

\section{Discovery of \thisMainObj}
\label{sec:discovery} 

DES \citealt{des_survey} is an optical survey targeting nearly 5000 square degrees of sky. It uses the Dark Energy Camera \citep[DECam,][]{2015AJ....150..150F} on the 4-meter Blanco telescope at the Cerro Tololo Inter-American Observatory in Chile. DECam is a prime-focus imager on Blanco with a 3 square degree field of view and a focal plane consisting of 62 $2K \times 4K$ red-sensitive science CCDs.
DES saw first light in 2012, and the nominal survey period of 520 nights over five years ran from August 2013 through February 2018. During this time, DES has operated in two survey modes. The Wide Survey observes the full of the survey area roughly twice per year in each of the \emph{grizY} bands. The Supernova Survey \citep{2012ApJ...753..152B} consists of ten 3 square degree regions that are observed roughly every 6 days in the \emph{griz} bands.
Due to the large survey area, high repetition, and deep limiting magnitude for single-epoch exposures ($r\sim$23.8 magnitude), DES has many applications in addition to its main cosmological objectives \citep{2016MNRAS.460.1270D}. It is well-suited for solar-system science, and in particular to the study of high-inclination populations. 

In this paper, we report the discovery of \thisMainObj, which has the largest semi-major axis of any 
object yet found by DES, and the highest inclination of any known extreme TNO. 
\thisMainObj\ was first detected at a heliocentric distance of 55~AU in the same set of observations from 2013-16 that were used to discover the dwarf planet candidate 2014~UZ$_{224}$ \citep{2017ApJ...839L..15G}.
The original detection of \thisMainObj\ came from a difference-imaging analysis of the wide field images \citep[using software from][]{2015AJ....150..172K}. Transient objects are found by image subtraction. Artifacts and low-quality detections are rejected using the techniques described in \citet{2015AJ....150...82G}. 
The surviving sources are compiled into a catalog of measurements, each of which corresponds to a transient at one epoch. 
From those, TNOs are extracted by identifying pairs of detections within 60 nights of each other whose angular separation is consistent with what would be expected for an object with perihelion $q>30$ AU given Earth's motion. 
These pairs are linked into chains of observations that correspond to the same object by testing the goodness of fit of the best-fit orbit for each chain. A reduced chi-squared $\chi^{2}/N < 2$ \citep{2000AJ....120.3323B} is considered a detection of a TNO. 

\begin{figure}[htbp] 
   \centering
   \includegraphics[width=3.4in]{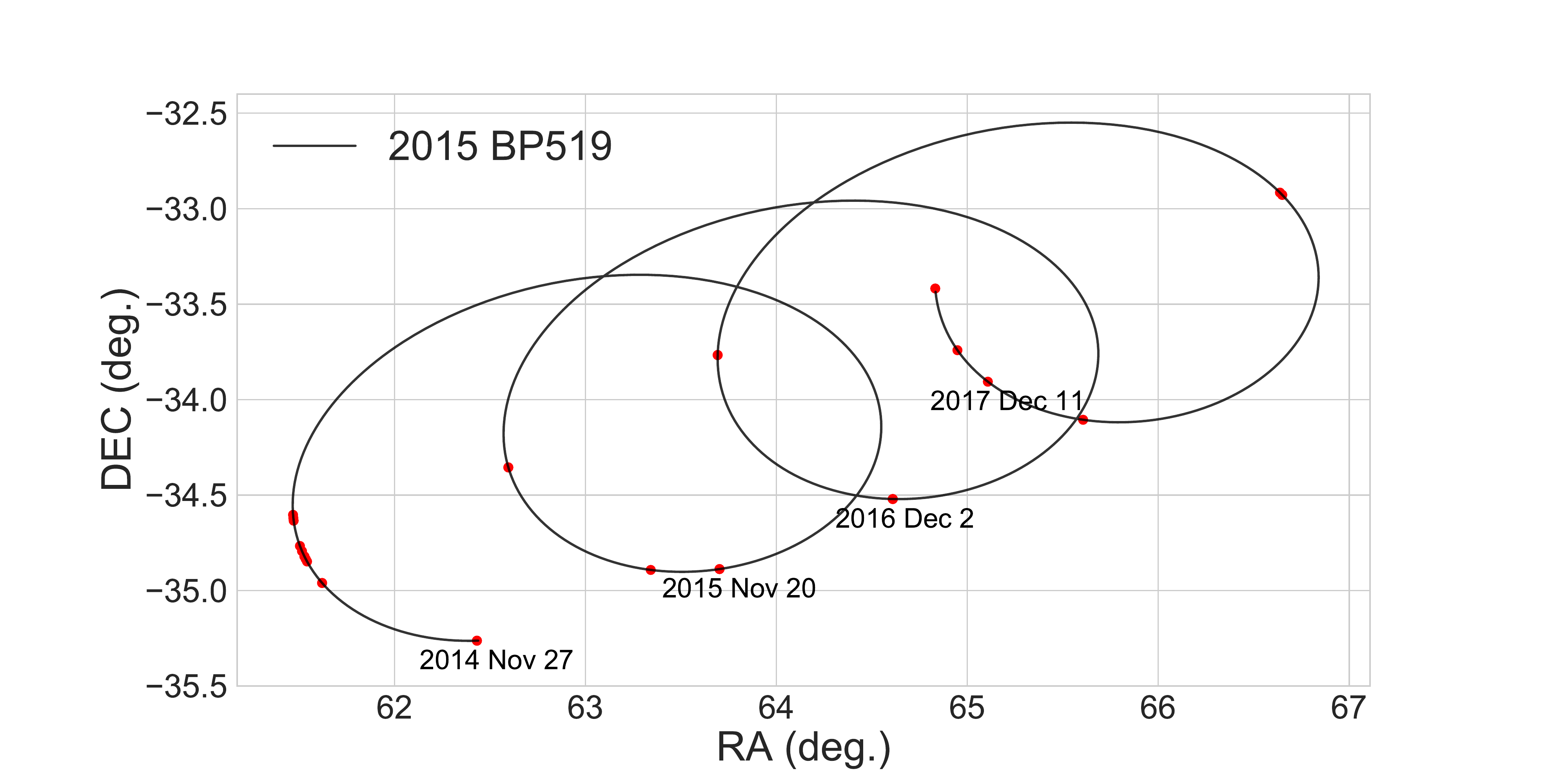} 
   \caption{
Trajectory of \thisMainObj\ over its measured four-opposition arc. Larger, red dots along the trajectory indicate points at which it was 
observed by DES.}
   \label{fig:discovery_arc}
\end{figure}

Although \thisMainObj\ was originally identified using data from observing campaigns 2-4, we have obtained additional observations  in two ways: first, some of the subsequent planned DES exposures provided additional serendipitous observations of this object. Second, we performed three targeted observations on 2 Feb. 2017 and 6-7 August 2017. The result is a series of 30 observations over four oppositions between 27 Nov. 2014 and 15 Feb. 2018, shown in Figure~\ref{fig:discovery_arc}. We computed astrometric positions using the \textsc{WCSfit} software described in \citet{2017PASP..129g4503B}, which provides astrometric solutions referenced to the Gaia DR1 catalog \citep{GAIA_DR1}. This includes corrections for
the effects of tree-ring and edge distortions on the DECam CCDs, as well as for chromatic terms from lateral color and differential atmospheric refraction. We obtain barycentric osculating orbital elements using the method of \citet{2000AJ....120.3323B}. For consistency with the orbital elements and uncertainties used in the simulation results presented below, our fit uses the 27 observations available through 11 Dec. 2017. The resulting fit has a $\chi^2$ of 48.2 for 48 degrees of freedom, and a mean residual of 29~mas. 
These orbital elements are shown in Table \ref{tab:orbital_elements}. 
\thisMainObj's inclination and orbital orientation relative to the other extreme TNOs is also visualized in Fig. \ref{fig:orbit3D} (where the orbital elements used for the other plotted extreme TNOs are given in the Appendix, Table \ref{tab:etno_table}). 
\begin{figure}[h]
\epsscale{1}
  \begin{center}
      \leavevmode
\includegraphics[width=3.4in]{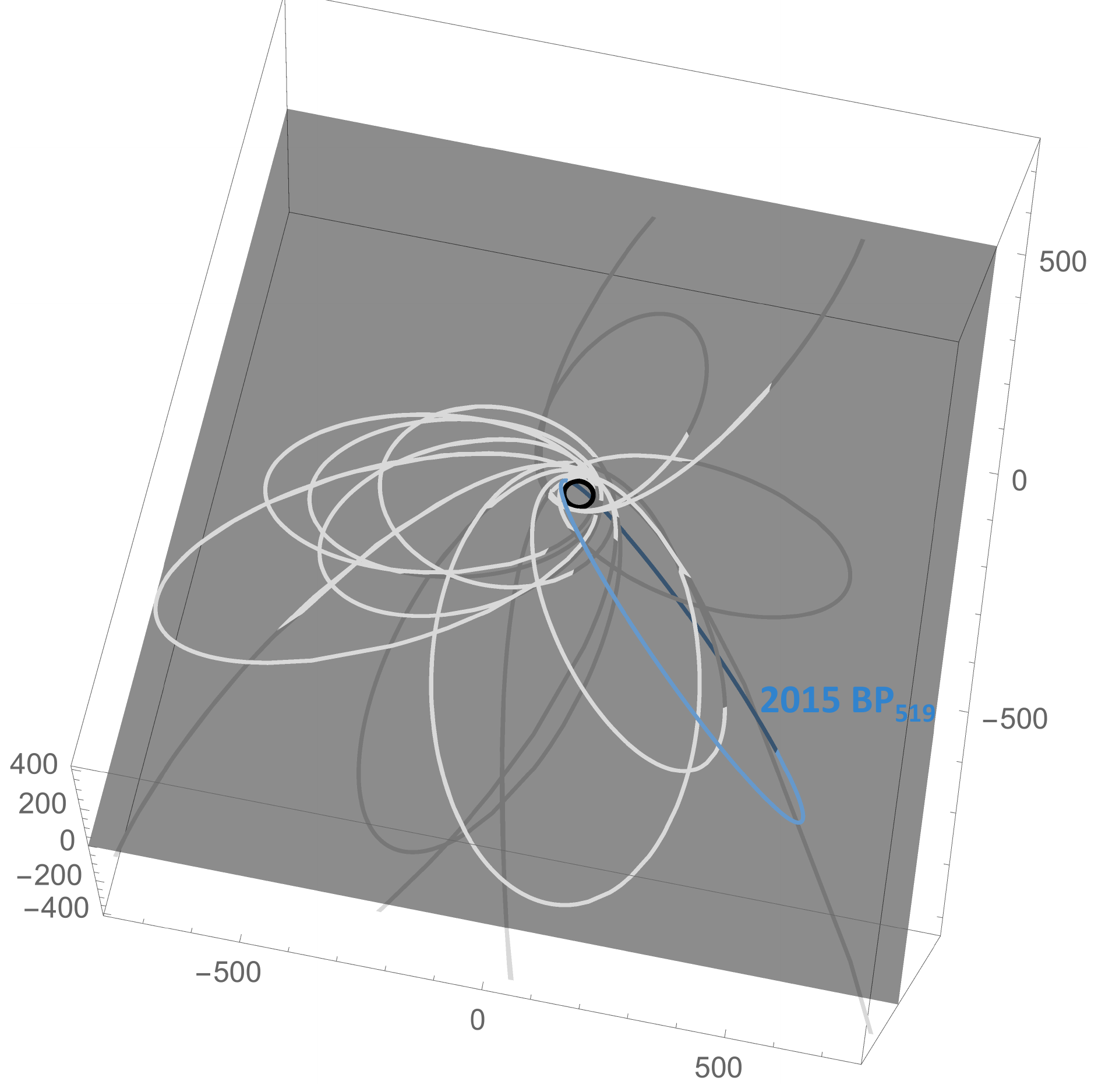}
\caption{A visual representation of the orbit of \thisMainObj, plotted with the other ETNOs as comparisons. For each orbit, the darker regions on the curve denote where an object falls below the plane of the solar system. \thisMainObj\ has the highest inclination of any extreme TNO discovered to date. The full, interactive 3D orbit visualization can be found at \url{https://smillholland.github.io/BP519/}.  }
\label{fig:orbit3D}
\end{center}
\end{figure}

These 27 observations of 2015 BP519 include 8 measurements in the \emph{g}-band, 9 in \emph{r}-band, 6 in \emph{i}-band, and 4 in \emph{z}-band. Few of these observations were taken in close temporal proximity. To compute the colors of this object, we therefore compute the corresponding absolute magnitude $H$ of each measurement to correct for the varying object-sun and object-earth distances as well as differences in observational phase. The \emph{g-r} color, for example, is then computed as $<H_g> - <H_r>$, and its uncertainty is $(<H_g^2>+<H_r^2>)^{1/2}$.
The moderately red \emph{g-r} and \emph{r-z} colors are consistent with the values measured in \citep{2017AJ....154..101P} for objects identified as dynamically excited.

For TNOs with magnitudes in the range $H\sim2-4$, measured visual albedos have been found to range between 0.07 and 0.21 \citep{2009Icar..201..284B,2013AA...557A..60L,2014ApJ...782..100F, 2016AJ....151...39G,2018ApJ...855L...6H}. With $H_r$ = 4.3, the diameter of \thisMainObj\ could range from 400-700 km, depending on whether the albedo falls near the high or low end of this range. 

Because the DES survey area lies predominantly out of the ecliptic, the status of \thisMainObj\ as the highest inclination TNO of those with semi-major axis $a>250$~AU and perihelion $q>30$~AU must be considered in the context of possible bias of the DES selection function.  To explore this issue, we simulate an ensemble of clones of \thisMainObj\ and test their recoverability in the DES TNO search pipeline. The orbital elements of these clones are drawn from the observed posteriors provided in Table \ref{tab:orbital_elements}, but with the inclination angle $i$ drawn from a uniform distribution between 0 and 180 degrees. We then compute the orbits of these objects and where the clones would fall on the nights DES observed. 

Using these synthetic orbital elements, we first remove any object that is not detectable by DES because it is either too faint
or outside the survey area. We then compute the position of each remaining clone at the time of every DES exposure belonging to the data set in which \thisMainObj\ was discovered, and determine which clones could be linked together into an orbit. The clones that could be identified as candidates are those with at least three observations on three different nights separated by less than 60 nights, and with observations on at least five different nights in total.

\begin{figure}[h]
\epsscale{1}
  \begin{center}
      \leavevmode
\includegraphics[width=3.4in]{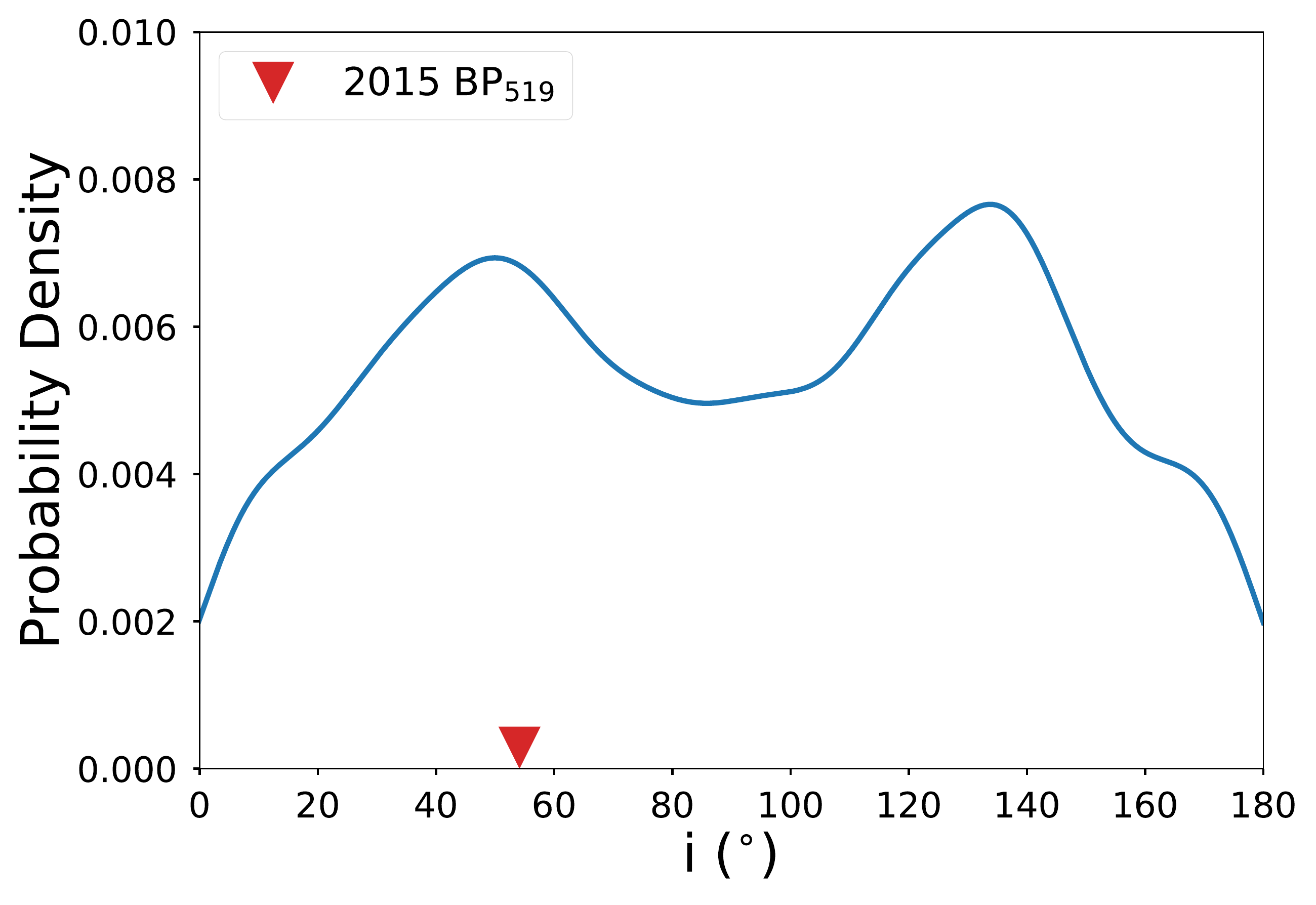}
\caption{The DES selection function for the discovery of objects with the orbital elements of \thisMainObj, but with varying inclination angles. The value for \thisMainObj\ is shown as the red triangle. The probability distribution is normalized so that the area under the curve is unity.}
\label{fig:bias}
\end{center}
\end{figure}

The distribution of clones that survives this process, and hence is potentially detectable, is presented in Fig. \ref{fig:bias}. This plot thus shows the sensitivity function for objects with the orbital parameters ($a,e,\omega,\Omega$) of \thisMainObj, but with varying orbital inclinations and mean anomalies. The resulting sensitivity function shows some structure, but is not heavily biased toward the observed inclination angle of \thisMainObj. 

\begin{figure}[htbp] 
\centering
\includegraphics[width=3.4in]{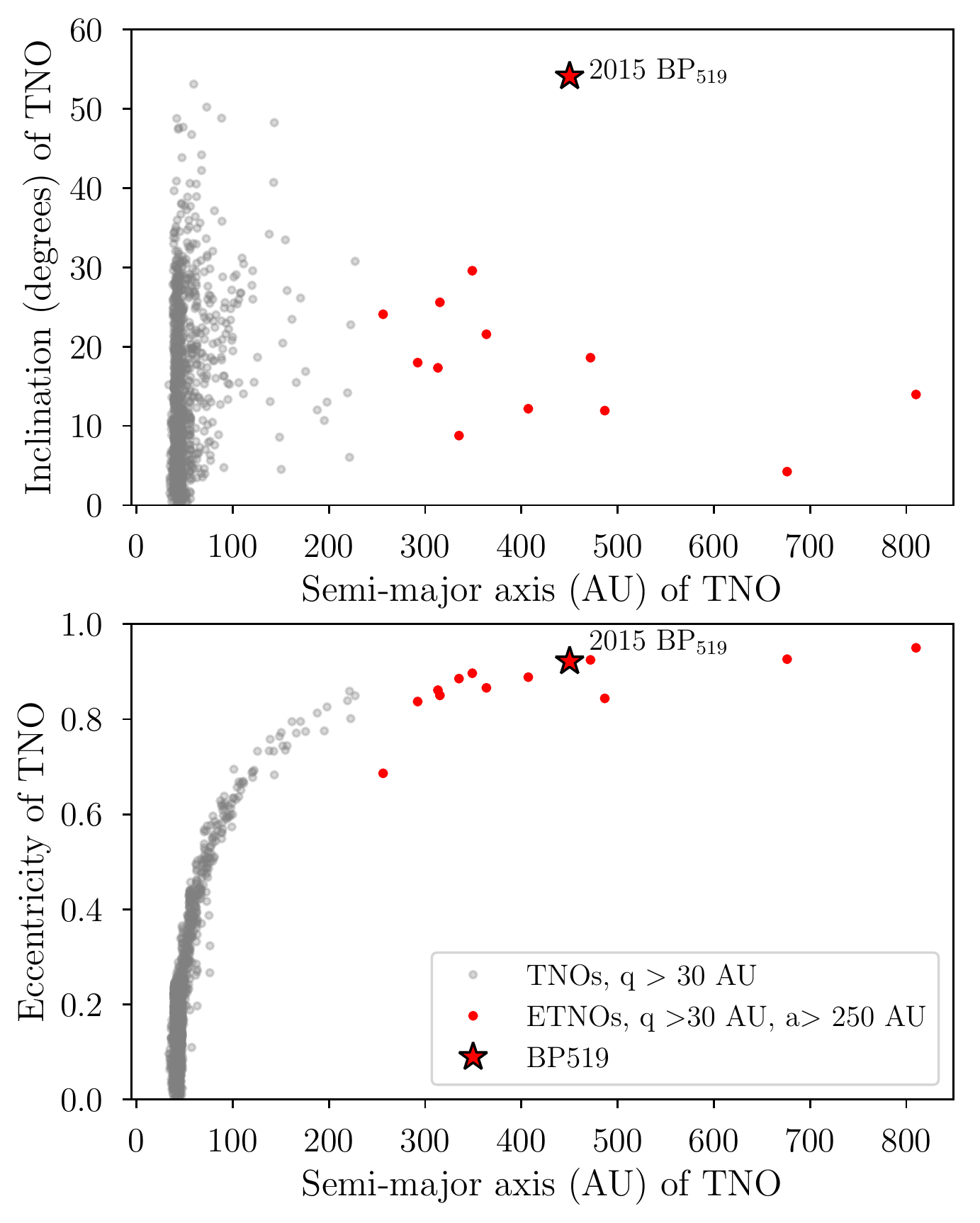}
\caption{The distributions of inclinations (top panel) and eccentricities (bottom panel) for the two populations of TNOs considered in this work: all objects with perihelia distances greater than 30 AU, and then the subset of those with semi-major axes greater than 250 AU. Orbits of known objects are fit from observations posted to the Minor Planet Center database. \thisMainObj\ represents the tail of the inclination distribution of the known TNOs, as well as the upper limit of eccentricities populated by TNOs.  }
\label{fig:hist_inclinations}
\end{figure}

\thisMainObj\ has the highest inclination of any known TNO (defined as objects with $q>30$ AU). \thisMainObj\ also has an extreme eccentricity (\thisMainObjECC). 
Figure \ref{fig:hist_inclinations} compares the inclination and eccentricity occurrences by semi-major axis of the regular and extreme ($a>$ 250 AU) TNO populations. Compared to the other known TNOs, \thisMainObj\ has the largest orbital inclination. Since the number of known ETNOs is small, however, is it unclear where \thisMainObj's inclination places it in the true distribution of ETNO inclinations. 
For regular TNOs (objects with perihelion distance $q>30$ AU but any semi-major axis), for which nearly 1500 have been discovered, \thisMainObj\ is the most extreme and seems to lie at the upper tail of the inclination distribution of known objects; among TNOs, \thisMainObj\ has the highest currently measured value, but this population is by no means complete. 

\begin{figure}[h]
\epsscale{1}
  \begin{center}
      \leavevmode
\includegraphics[width=3.4in]{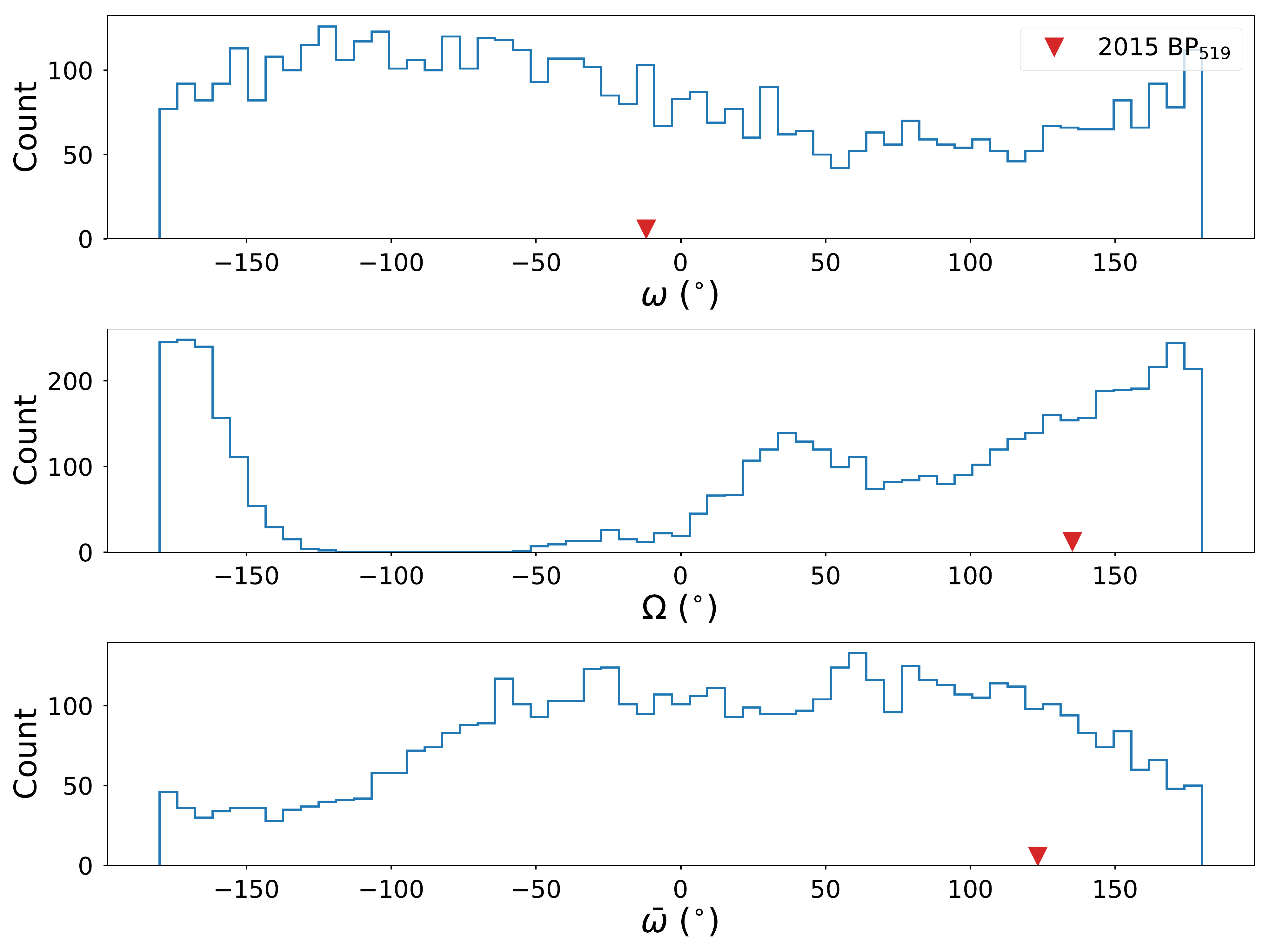}
\caption{The DES selection function for the discovery of objects with the orbital elements ($a,e,i$) of \thisMainObj, but with varying angles $\omega$, $\Omega$. The observed values for \thisMainObj\ are shown as red triangles on each panel. For objects with the orbital elements ($a,e,i$) of \thisMainObj, the DES observation bias allows discovery of $\omega$ and $\Omega$ subtending most of the allowable ranges.}
\label{fig:bias2}
\end{center}
\end{figure}

In Fig. \ref{fig:bias2}, we plot a sensitivity histogram computed in the method described above, but for objects with the orbital parameters ($a,e,i$) of \thisMainObj, and varying $\omega$ and $\Omega$. As was true for the previous sensitivity function, the final sensitivity histogram shows some structure in each orbital angle of interest, but is not heavily biased towards the measured angles of \thisMainObj\ (which happen to be consistent with the region of clustering that was used to predict Planet Nine, as will be discussed further in later sections of this paper; \citealt{st14, bb16}). 

\begin{deluxetable}{lc}
\tablecaption{Orbital Elements of \thisMainObj}
\tablewidth{0pt}
\tablehead{
  \colhead{Parameter } &
  \colhead{Value}   \cr
}
\startdata
a & 448.99   $\pm$ 0.49 AU \\
e & 0.92149 $\pm$ 0.00009\\
i& 54.1107 $\pm$ 0.00001 deg\\
$\omega$ & 348.058 $\pm$ 0.00136 deg\\
$\Omega$ & 135.2131  $\pm$ 0.00010 deg\\
Time of Perihelion (JD) & 2473015.55 $\pm$ 0.56  \\
Perihelion & 35.249 $\pm$ 0.078 AU\\
Aphelion& 862.733 $\pm$ 0.972 AU\\
Orbital Period& 9513.84 $\pm$ 15.42 years\\
Absolute magnitude & $H_r = 4.3$\\
\emph{g-r} (mag) &  0.79 $\pm$ 0.17\\
\emph{r-i} (mag) &   0.19 $\pm$ 0.12\\
\emph{r-z} (mag) &   0.42 $\pm$ 0.15\\
\emph{i-z} (mag) &   0.23 $\pm$ 0.15\\
\enddata
\tablecomments{\thisMainObj\ barycentric osculating elements at epoch 2456988.83, based on 27 observations over a 1110-day arc from 27 Nov.
2014 to 12 Dec. 2017. \thisMainObj\ has a mean anomaly 358.34 degrees and will reach perihelion on 14 Oct. 2058.} 
\label{tab:orbital_elements}
\end{deluxetable}

\section{Characterization of \thisMainObj}
\label{sec:characterization}

As a starting point, we consider the dynamical behavior of \thisMainObj\ using a secular treatment. The basic approach is outlined and compared with numerical N-body experiments in subsection 
\ref{sec:secular}, and this formalism is used to elucidate the extreme nature of this object in subsection \ref{sec:extreme}.

\subsection{Secular Dynamics}
\label{sec:secular} 

A secular approach averages over the mean motion of solar system objects and thus allows for a simplified treatment of the long-term dynamics. \citet{1962AJ.....67..591K} provided secular equations for the orbital evolution of small bodies with high inclinations and eccentricities in the presence of an inner perturber. Here we want to describe the behavior of \thisMainObj, which orbits outside a system of four interior perturbers (namely, the known giant planets). The contribution from the terrestrial planets is negligible in this context. We can write the mean perturbing function $R_{m}$ for a test particle evolving in the presence of a set of inner planets in the form 
\be
\begin{split}
R_m = \frac{G}{16 a^3 \left( 1-e^2 \right)^{3/2}} \Biggl[ \quantityC  (1+3 \cos 2i) \\ + 
\frac{9  \quantityE   \left( 2+3 e^2 \right) (9+20 \cos 2i +35 \cos4i)} {512 a^2 \left( 1-e^2 \right) ^{2}} \\
+ \frac{9  \quantityE \ 40 e^2 (5+7 \cos2i) \cos 2\omega
\sin^2 i }{512 a^2 \left( 1-e^2 \right) ^{2}} \Biggr]\,,
\label{eq:distfunct}
\end{split}
\ee
where the effects of the inner planets are included here as a mean moment of inertia \citep{2012Icar..220..392G}. In this expression, $G$ is the gravitational constant ($G= 4\pi^{2}$; we work in units of solar mass, AU, and year), $(a,e,i)$ are the orbital elements of the test particle, $\quantityC$ is the moment of inertia in the direction out of the plane containing the giant planets, and the label $j$ denotes each giant planet under consideration.

\begin{figure}[htbp] 
   \centering
   \includegraphics[width=3.4in]{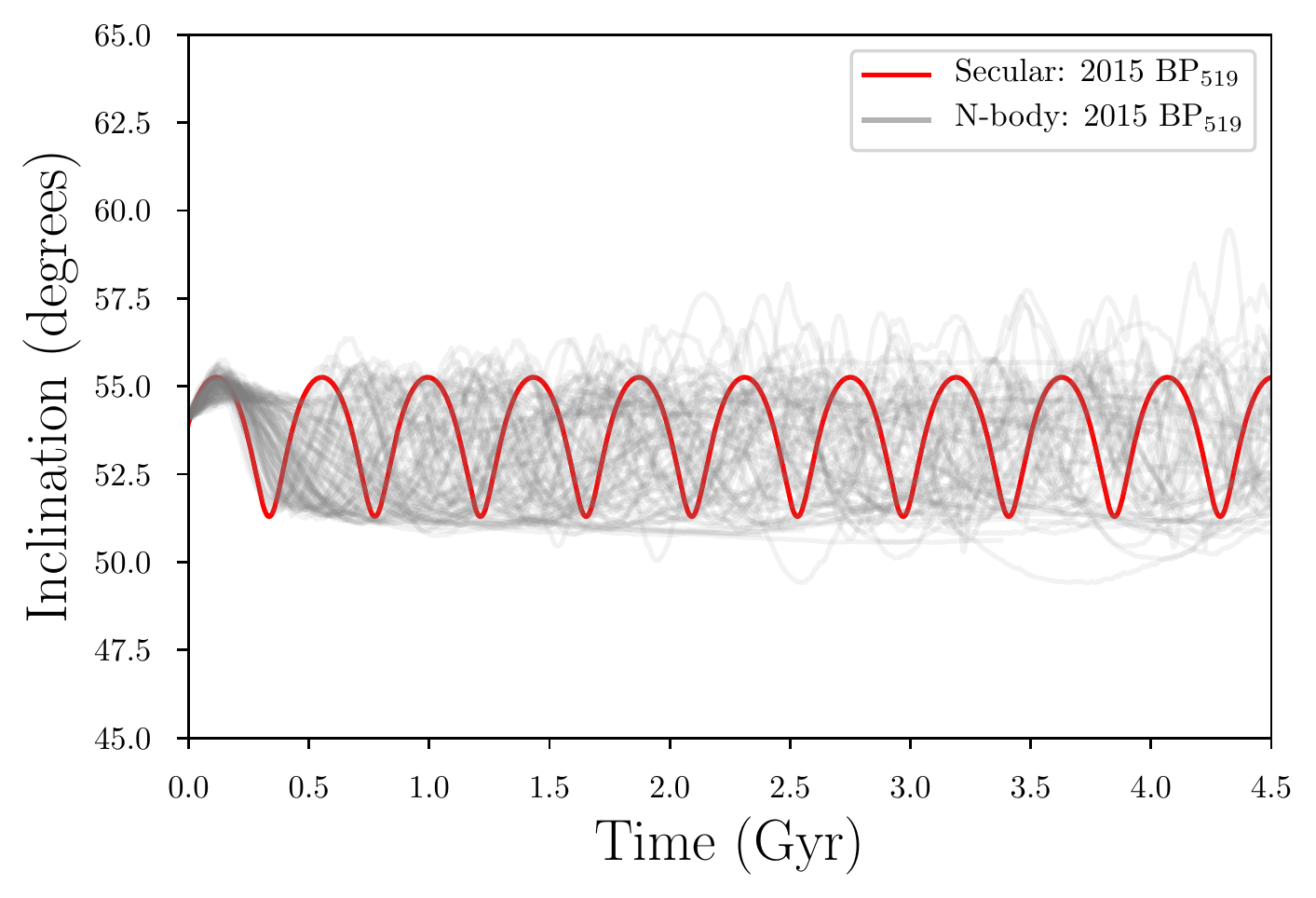} 
   \caption{
The future evolution of \thisMainObj\ (using its current day best-fit orbital parameters as initial conditions) in the presence of the known solar system. The secular curve plotted as a solid dark line was solved from the disturbing function (Equation \ref{eq:distfunct}) and the best-fit orbital elements of \thisMainObj. The numerical results, plotted as grey lines, are drawn from simulation Set 1, where the orbit of \thisMainObj\ is evolved in the presence of the known solar system for 4.5 Gyr. See Table \ref{tab:sim_sets_params} for more details on the simulation parameters. } 
   \label{fig:secular_plot}
\end{figure}

From this secular Hamiltonian, we can derive an expression for the time evolution of the inclination angle using Lagrange's planetary equations, which takes the form 
\be
\begin{split}
\frac{di}{dt} = -\tan{(i/2)} (na^{2} \sqrt{1-e^{2}} \ )^{-1} \left(\frac{dR_m}{d\epsilon}+ \frac{dR_m}{d\varpi} \right) \\- (na^{2} \sqrt{1-e^{2}} \sin{i} )^{-1} \frac{dR_m}{d\Omega}\,.
\end{split}
\ee
where $n =(GM / a^{3})^{1/2}$ is the mean motion, $M$ is the mass of the central body, and $\epsilon$ is the mean longitude at epoch.
Combining the previous two expressions yields
the equation of motion for $i$: 
\be
\frac{di}{dt} = \frac{45e^{2} G^{1/2} \quantityE}{1024\ M^{1/2} \ a^{11/2} (1-e^{2})^{4}} (5 + 7 \cos{2i}) \sin{2i} \sin{2 \omega}\,.
\label{eq:precess}
\ee
Analogous equations can also be constructed using the other Lagrange planetary equations, resulting in equations of motion for $de/dt$, $d\omega/dt$, $d\Omega/dt$ (see Equations 7-11 of \citealt{2012Icar..220..392G}), and $da/dt = 0$. Using the known (estimated) orbital elements for \thisMainObj\ (see Table \ref{tab:orbital_elements}) as initial conditions, we simultaneously solved these five equations of motion, resulting in predicted secular evolution for \thisMainObj's orbital evolution.
This result is shown in Fig. \ref{fig:secular_plot} as the solid curve. The figure shows additional curves in thinner grey lines corresponding to the orbital evolution computed for the same initial conditions, but using full N-body integrations instead of secular theory. These simulations are described in full in the following section, and their parameters are also summarized as Set 1 in Table \ref{tab:sim_sets_params}. As a quick summary, these integrations are computed using the \texttt{Mercury6} integration package \citep{m6}, using the hybrid symplectic and Bulirsch-Stoer (B-S) integrator and a time-step of 20 days. In these simulations, all four of the known giant planets are treated as active bodies (rather than being modeled using the $J_2$ approximation that is often used).

Figure \ref{fig:secular_plot} shows that the secular approximation provides a good order of magnitude description of the time evolution of the inclination angle, even through the secular approximation does not include the scattering interactions that lead to slight divergence in the N-body simulations. 
Both the secular and N-body treatments predict that, in the known solar system, the inclination of \thisMainObj\ will remain fairly well constrained around its presently observed value.

\subsection{The Extreme Nature of \thisMainObj}
\label{sec:extreme} 
\begin{figure}[htbp] 
   \centering
   \includegraphics[width=3.4in]{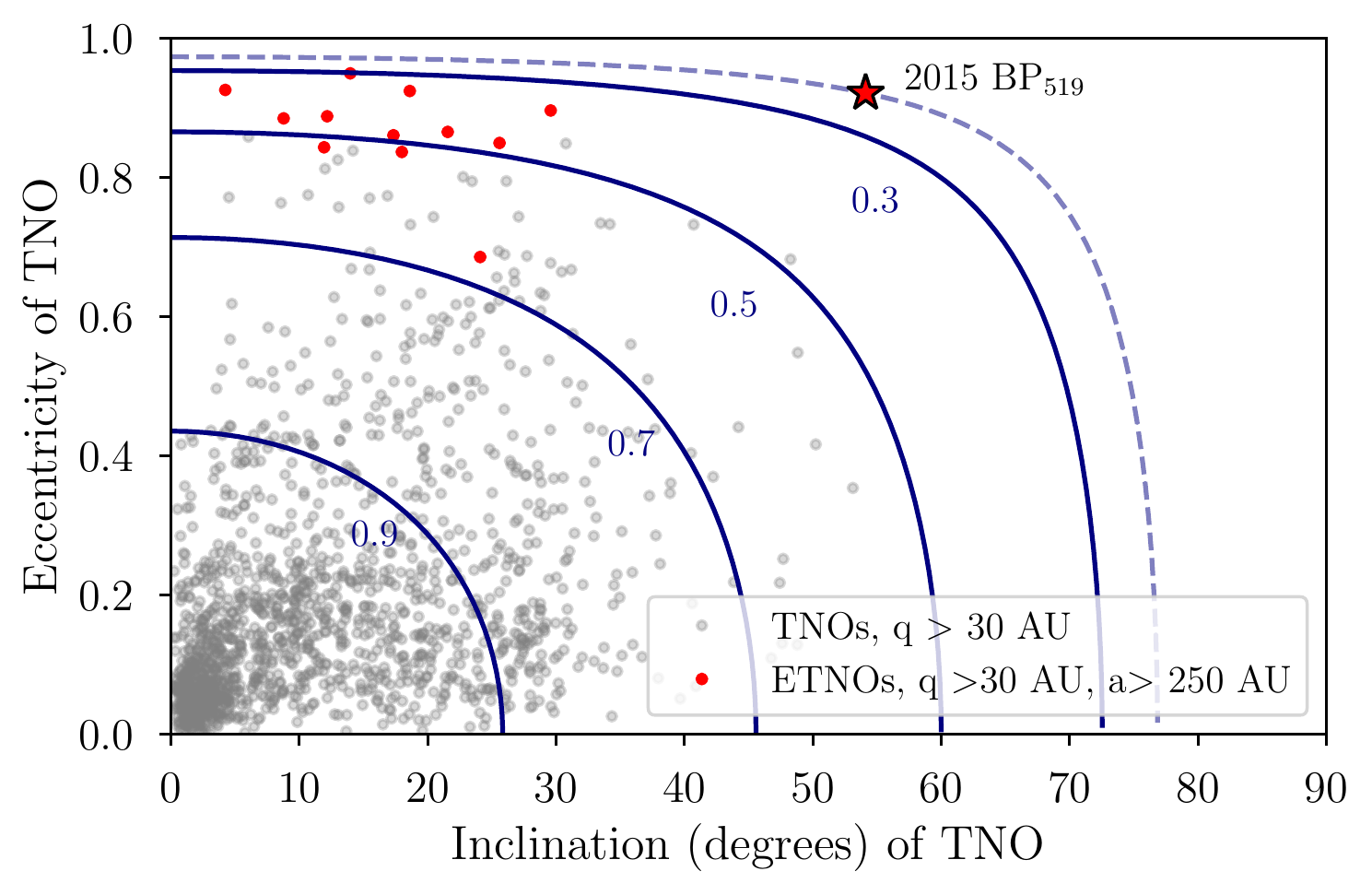} 
   \caption{
We plot curves of constant $\eta_{0}$ in ($i,e$) space (see Equation \ref{eq:eta}). Also plotted are (in grey) the  orbital elements of all objects with perihelion distances outside of Neptune and data quality flags of 6 or better, as reported to the Minor Planet Center database (files downloaded 10/25/2017) and (in red) the subset of those objects that also have a semi-major axis measured to be $a>250$ AU. With $\eta_{0} = 0.2274$, \thisMainObj\ has the lowest value of $\eta_{0}$ out of any TNO with $q> 30$ AU that has been discovered thus far. This metric, which measures the extremeness of the ($i, e$) of each object, characterizes \thisMainObj\ as the most extreme of the extreme TNOs. }
   \label{fig:eta_level_curves}
\end{figure}

Although the orbit of \thisMainObj\ is highly unusual among known TNOs, we need a quantitative assessment of its properties relative to other TNOs of its dynamical class. Toward that end, we consider the Kozai Hamiltonian written in Delaunay coordinates \citep{1996CeMDA..64..209T}, for which the action $H$ is defined as:
\be
H = \sqrt{a (1-e^{2})} \cos{i}\,.
\ee
Note that this action is equivalent to the standard $`{\cal H}'$ variable in Delaunay coordinates \citep{md1999} and is a constant of the motion in the quadrupolar approximation. The action $L=\sqrt{a}$ will also be constant, as the Kozai Hamiltonian averaged over the mean anomaly and thus rotationally invariant and thus depends only on action $G = \sqrt{a(1-e^{2})}$ and coordinate $g=\omega$, with actions $L$ and $H$ being conserved. 
Next, we define a reduced Kozai action $\eta_0$, which has the form 
\be
\eta_{0} = \sqrt{(1-e^{2})} \cos{i}\,.
\label{eq:eta}
\ee
As action $L$ is conserved for the Kozai Hamiltonian, this reduced form of action $H$ should also be conserved.
Note that $\eta_{0}$ is the specific angular momentum vector in the direction out of the plane of the solar system (\citealt{1999CeMDA..75..125K}; we follow the notation in \citealt{2016CeMDA.tmp...31S}). 
For sufficiently distant TNOs, the potential of the solar system is effectively axially symmetric (but not spherically symmetric), so that the $z$-component of angular momentum (but not total angular momentum) is conserved. In the known solar system, TNOs with constant semi-major axis are thus expected to evolve in $(i,e)$ space along contours of constant $\eta_{0}$. In Fig. \ref{fig:eta_level_curves}, we overlay curves of constant $\eta_{0}$ on a plot comparing the $i,e$ of all TNOs and ETNOs discovered so far. Compared to previously discovered objects, \thisMainObj\ has the lowest $\eta_{0}$ value which signifies its relative extremeness. 

\vspace{10mm}
\section{Full Dynamics of \thisMainObj\ \\ in the Known Solar System }

\label{sec:numericalknown} 

The analytic formulation presented in Section \ref{sec:characterization} classifies \thisMainObj\ as the most extreme TNO discovered in the outer solar system to date, due to its high inclination, high eccentricity, and large semi-major axis. 
However, the secular approximation used in the previous section neglects the importance of interactions with Neptune, which will occur when \thisMainObj\ reaches its perihelion. 
\thisMainObj's relatively small perihelion distance ($\approx 35$ AU) suggests that it will be subject to repeated strong interactions with Neptune, which will change the energy of \thisMainObj's orbit by a factor of roughly $6\times10^{-6}$ per perihelion crossing \citep[when this process can be modeled as a random walk; see Fig. 1 of][]{1987AJ.....94.1330D}. 
The change in orbital energy will also lead to change in the semi-major axis of the orbit, and as a result, the level curves presented in Fig. \ref{fig:eta_level_curves} may not truly represent the evolution of \thisMainObj\ over extended spans of time. Instead, quantities which appear as constants of motion in the previous section ($\eta_{0}$) will no longer be conserved as \thisMainObj\ changes its 
orbital elements, in particular its semi major axis,  due to interactions with Neptune.
The true orbital evolution, being the result of a chaotic process, will also vary widely between trials in numerical integrations. 
As shown in Fig. \ref{fig:secular_plot}, the numerically computed orbital evolution does not perfectly match the secular expectation, and multiple integrations of the same object will give slightly different periods and amplitudes of evolution.

To fully test the effect of additional dynamics not encapsulated by the secular analysis of the previous section, we perform a suite of numerical N-body simulations using computing resources provided by Open Science Grid \citep{osg1, osg2, osg3} through the Extreme Science and Engineering Discovery Environment (XSEDE) portal \citep{xsede}.
These simulations include the new body \thisMainObj\ and all of the relevant known solar system objects (the case of Planet Nine is considered in the following section, but is excluded from this initial set of simulations).

\subsection{Numerical Evolution of \thisMainObj\ \\ in the Known Solar System}
\label{sec:numnop9}

The precession time scales and orbital evolution of \thisMainObj\ can be tested more directly with numerical N-body simulations. To examine the complete evolution of \thisMainObj\ in the known solar system, we perform a suite of numerical integrations using the \texttt{Mercury6} integration package \citep{m6}. 
We exclude the terrestrial planets from the simulations, but include the gas giants (Jupiter, Saturn, Uranus, and Neptune) as active, massive particles with their currently measured masses and orbital elements. 
We start with a time-step of 20 days, which is roughly 0.5\% of Jupiter's orbital period. 
We use the hybrid symplectic and Bulirsch-Stoer (B-S) integrator built into \texttt{Mercury6} and conserve energy to better than 1 part in $10^{9}$ over the course of the 4.5 Gyr integrations. 
The orbital elements for \thisMainObj\ are drawn from the covariance matrix derived from the fit to the DES data. 
Fifty-two simulations are run of the solar system, each with five clones of \thisMainObj. Half of these simulations are integrated forward in time from the current day, and the other half evolve back in time for 4.5 Gyr.
Other parameters used for this set of simulations (which we call Set 1 in this work) are given in Table \ref{tab:sim_sets_params}. 

\begin{deluxetable*}{lccccccc}
\tablecaption{Simulation Sets used in this work}
\tablewidth{0pt}
\tablehead{
  \colhead{Set} &
  \colhead{Initial Timestep}  &
  \colhead{Active Planets} &
  \colhead{$J_{2}$}     &
  \colhead{Abs. Radius }   &
  \colhead{Backwards Clones}   &
  \colhead{Forward Clones}   &
  \colhead{Details}   \\
}  
\startdata
Set 1	&	20 days	&	4 (JSUN)	&	2$\times10^{-7}$ 	& 4.65$\times10^{-3}$  AU & 130	&	130	&	No P9 	\\
Set 2	&	3000 days 	&	1 (P9)	&	0.00015244 	& 20 AU & 0 & 1000 &  low $i^{1}$, P9$^{2}$  					\\
Set 3	&	3000 days 	&	1 (P9)	&	0.00015244 	& 20 AU & 0 & 1000 &  P9$^{2}$  					\\
Set 4	&	200 days 	&	2 (N,P9)	&	0.00036247 	& 9.8 AU & 0 & 600	& P9$^{2}$		\\
Set 5	&	20 days	&	5 (JSUN, P9)	&	2$\times10^{-7}$	& 4.65$\times10^{-3}$  AU & 0	&	130	& P9$^{2}$  \\
	\enddata
\tablecomments{A list of the sets of simulations used in this work, with their relevant parameters listed. When included as active particles in a simulation, gas giant planets are denoted by their first initials (J for Jupiter, S for Saturn, U for Uranus, N for Neptune) and when Planet Nine is included, it is denoted by P9. The absorbing radius is the radius of the central body in the simulations. The ejection radius is set to 10000 AU for all simulation sets, and all integrations are run for 4.5 Gyr. Except when denoted otherwise, the orbital elements of \thisMainObj\ are drawn from the covariance matrix describing the best-fit values and errors in Table \ref{tab:orbital_elements}. Simulations were run in batches; for simulation Sets 1 and 5, 5 clones of \thisMainObj\ were included as test particles in each individual integration. For simulation Sets 2, 3, and 4, 10 clones were included in each integration. $^{1}$ Inclination of \thisMainObj\ was drawn from a half-normal distribution around 0 degrees with a width of 5 degrees. $^{2}$ The best-fit version of Planet Nine from \citet{2017AJ....154..229B} was used (700 AU, 0.6 eccentricity, 20 degrees inclination). The solar quadrupole moment $J_{2}$ is defined by Eq. \ref{eq:j2} when any giant planets are absorbed, and set to the solar value otherwise \citep{2003ApSS.284.1159P}.      } 
\label{tab:sim_sets_params}
\end{deluxetable*}

\begin{figure*}[htbp] 
\centering
\includegraphics[width=7in]{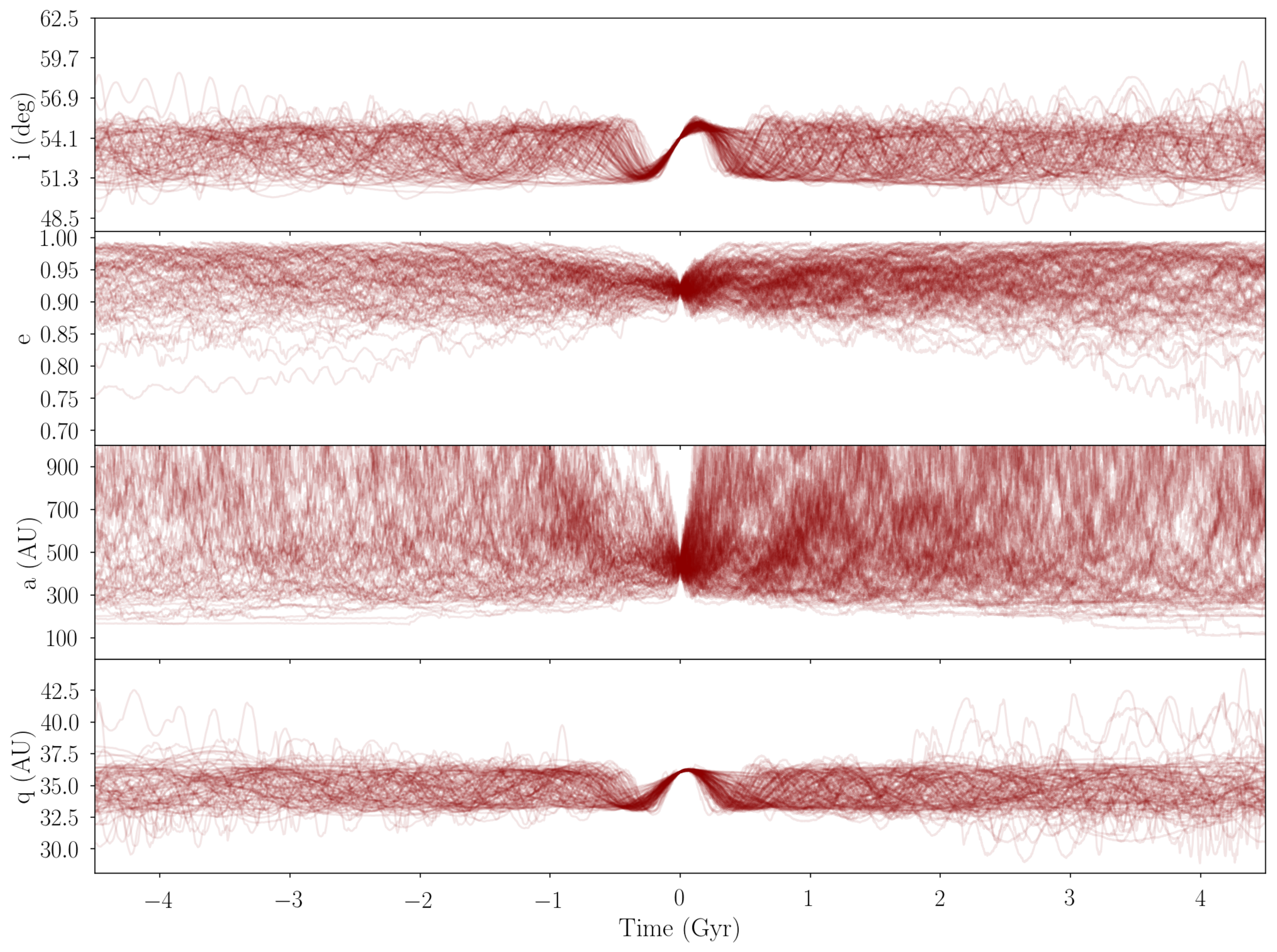}
\caption{The results of numerical simulations where \thisMainObj\ is evolved forward and backward in time in the presence of the four giant planets (Set 1; see Table \ref{tab:sim_sets_params}). All trials are plotted here; curves that end prematurely before 4.5 Gyr correspond to the integrations where a clone becomes dynamically unstable (collision into the central body, ejection from the system, physical collision with a planet, or a scattering event resulting in an unbound orbit). For trials that remain dynamically stable, the inclination and eccentricity are relatively well constrained to values near their initial conditions. The semi-major axis of \thisMainObj\ diffuses rapidly. }
\label{fig:backforth}
\end{figure*}
The results of these integrations are presented in Fig. \ref{fig:backforth}, and demonstrate that the semi-major axis of \thisMainObj\ diffuses widely in the presence of the known giant planets.
The perihelion distance tends to remain fairly well-confined near the initial value of even as the orbital energy changes due to repeated kicks from Neptune.

The Hamiltonian used in Section \ref{sec:extreme} requires that the semi-major axis of the particle remain roughly constant. From these simulations, it is clear that the semi-major axis of \thisMainObj\ tends to change significantly over relatively rapid ($\sim10^{6}-10^{7}$ year) timescales. 
As such, Equation \ref{eq:eta} is a good model for \thisMainObj's short-term dynamical behavior, but not its long-term orbital evolution. 

\subsection{Generating Highly Inclined Objects \\ in the Known Solar System}
\label{sec:highly_inclined}
In the previous section, we used numerical simulations to determine the expected evolution of \thisMainObj\ in the presence of the four known gas giants. The results show that the inclination of \thisMainObj\ tends to be confined to within a range of roughly 5 degrees. Although the semi-major axis diffuses over a wide range of values, the corresponding evolution in eccentricity is constrained by the behavior of the perihelion: to leading order, the perihelion distance of \thisMainObj\ remains well confined. Specifically, $q$ is constant to within $\sim$5 AU over the entire envelope of all dynamically stable clones. The eccentricity evolution of \thisMainObj\ is thus explained by the requirement that the perihelion remains nearly constant as the semi-major axis varies. Moreover, this behavior is mediated by Neptune. The orbital evolution is consistent with that expected for a member of the scattered disk.  

The high present-day inclination of \thisMainObj\ is more difficult to explain. In the numerical simulations shown in Figure \ref{fig:backforth} in the context of the currently-observed solar system, the orbital inclination of \thisMainObj\ is found to remain roughly constant. This trend holds for simulations running both backwards and forwards in time. Since the solar system formed from a disk, we expect the orbital inclination of \thisMainObj\ to be low at birth. The transition from an initially low inclination orbit to the present-day (high) value must be explained by some mechanism that is not included in our simulations. Some possible explanations include the following: a passing star could excite objects to highly inclined and eccentric orbits; a particularly favorable impact parameter during a close encounter with Neptune could excite an object out of the plane of the solar system; the high inclination could be a fossil from violent migration processes in the early solar system; the self-gravity of a large disk of planetesimals in the scattered disk; and finally, the existence of proposed solar system member Planet Nine, which could lead to secular evolution in eccentricity and inclination for long-period TNOs, thereby producing the current-day orbit of \thisMainObj. In this section, we briefly consider the first three possibilities, and then examine the Planet Nine hypothesis in detail in Section \ref{sec:planetnine}. 

\subsubsection{Scattering interactions with other stars} 

As most planetary systems form in clusters
\citep{ladalada,porras}, the solar system is likely to have formed in such an environment \citep{adams2010}. Dynamical interactions between cluster members can shape the dynamics of the constituent planetary systems \citep{2012Icar..217....1B}. The interactions tend to have a moderate effect \citep{adams2006}, but can nonetheless sculpt the outer portions of the planetary system or the original disk, which are of interest here. 
Interactions in the birth cluster are expected to dominate over those that occur later on in the field, but the latter can still be significant. 
If the trajectory of a binary or single star brings it sufficiently close to a star hosting a planetary system, the geometry of the planetary system can be altered  \citep{2011MNRAS.418.1272J}. For example, 
\citet{2004Natur.432..598K} discuss the possibility that Sedna's orbit is the result of a passing star perturbing the orbit of objects in the Kuiper Belt. They find that if such a star had its own disk of planets and planetesimals, then some objects could be captured into our solar system onto high-inclination orbits. It is thus possible that \thisMainObj\ is the result of interactions between our solar system and another external, perturbing body. The interaction cross sections for such events are much larger at the low fly-by speed realized in young embedded clusters \citep{gdawgold}, so that the required event is more likely to occur in the birth cluster (compared to the field). 


\subsubsection{Scattering interactions with Neptune} 

As discussed in \citet{1987AJ.....94.1330D}, TNOs will experience a perturbation in orbital energy at each periapsis, when the TNO passes closest to the orbit of Neptune. Although Fig. \ref{fig:backforth} demonstrates that in our set of backwards integrations \thisMainObj\ has retained \emph{roughly} the same inclination for the past 4.5 Gyr, there is some variation among the individual trials. More specifically, one particular integration in the backwards time direction attained (at one point) an orbital inclination of 60 degrees, although such a large value was not attained in any of the other integrations in either direction. With a large enough set of simulations, one could find the probability that \thisMainObj\ could originate in an orbit closer to the plane of the solar system, and subsequently evolve into its present orbit. In this scenario, \thisMainObj\ could have reached its high inclination from a series of extreme 
scattering events with Neptune. Our current set of numerical simulations shows that this scenario is possible, but unlikely. 

\subsubsection{Remnant of planetary migration}

The Nice model \citep{2005Natur.435..459T, 2005Natur.435..462M,2005Natur.435..466G} suggests that even if the solar system starts as a roughly co-planar disk, the planets attain their small eccentricities and inclinations through scattering events with the large reservoir of planetesimals in the outer solar system. Some of these bodies will be forced to high eccentricities and inclinations, while others will be able to maintain their lower $(e,i)$ distributions \citep{2008Icar..196..258L}. This scenario is characterized by a short period of extreme instability, which corresponds to the Late Heavy Bombardment inferred in the history of our solar system (at an age of $\sim$600 Myr). As a result of this violent period, high-inclination objects can be created from objects originating at the outer edge of the planetesimal disk.  Although it is unclear how an object with a semi-major axis as high as that of \thisMainObj\ would be generated in this process, we cannot exclude the idea that \thisMainObj's currently observed orbital inclination may come from a period of violent instability in the early history of the solar system.

Another explanation for high semi-major axis, high eccentricity orbits could be the diffusion hypothesis proposed in \citet{2017AJ....153..262B} for the generation of 2013 SY$_{99}$'s orbit. Objects with the longest orbital periods may sequentially scatter outwards, detach their perihelia through galactic tides, and then diffuse inwards into orbits with long periods and detached perihelia. Galactic tides start to dominate once an object attains a semi-major axis of roughly 3000 AU or more \citep{1987AJ.....94.1330D}, meaning that the currently-observed TNOs are not generally susceptible to these effects. This mechanism does appear to describe 2013 SY$_{99}$, an object with a semi-major axis $a \approx 730$ AU and an eccentricity of 0.93, which fits into the dynamical class of objects that would be produced by this mechanism. However, \thisMainObj's perihelion is not sufficiently detached (35 AU vs. 50 AU for 2013 SY$_{99}$) for this mechanism to operate. 

Another explanation for this object's extreme orbit could be galactic tides acting on remnants of the inner Oort cloud. It has been suggested \citep{2012MNRAS.420.3396B} that centaurs may come from the inner Oort cloud rather than the scattered disk. \citet{2012Icar..217....1B} shows that the median inclination of the inner Oort cloud should be around 50 degrees. As mentioned in \citet{2012Icar..217....1B}, the number and orbital parameters of objects with large semi-major axis can be used to constrain birth cluster properties. An object at 450 AU would be near the inner 2-5\% of the cloud, depending on the density profile used. However, objects formed via this mechanism (such as SY99, \citealt{2017AJ....153..262B}) would be expected to have detached perihelia distances, which \thisMainObj\ does not. 

\citet{2018AJ....155...75S} discuss the possibility that a potentially planetary-mass object (sub-earth mass) could have formed among the giant planets, and its influence during Neptune's migration could have excited TNOs to present-day high inclinations. This object is distinct from the Planet Nine discussed in the next section.

\subsubsection{Self-gravity of the scattered disk}
A sufficiently large (1-10 Earth masses in total mass), eccentric disk would experience an instability due to the self-gravity of the disk \citep{2016MNRAS.457L..89M}. This proposed instability could cause clustering in $\omega$ (as observed) for the objects experiencing the instability, and a subsequent pumping of inclination for objects that find their apocenter above the orbital plane \citep{2018arXiv180503651M}. This would result in the population of high inclinations for eccentric objects. \thisMainObj\ could undergo this mechanism if the scattered disk contains enough mass to cause the instability: for this explanation to be feasible, a large number of additional objects in the scattered disk will need to be found, as the early mass of the scattered disk must have been high for this instability to occur.  

\section{Dynamics in the Presence of Planet Nine}
\label{sec:planetnine} 

\begin{figure}[htbp] 
\centering
\includegraphics[width=3.4in]{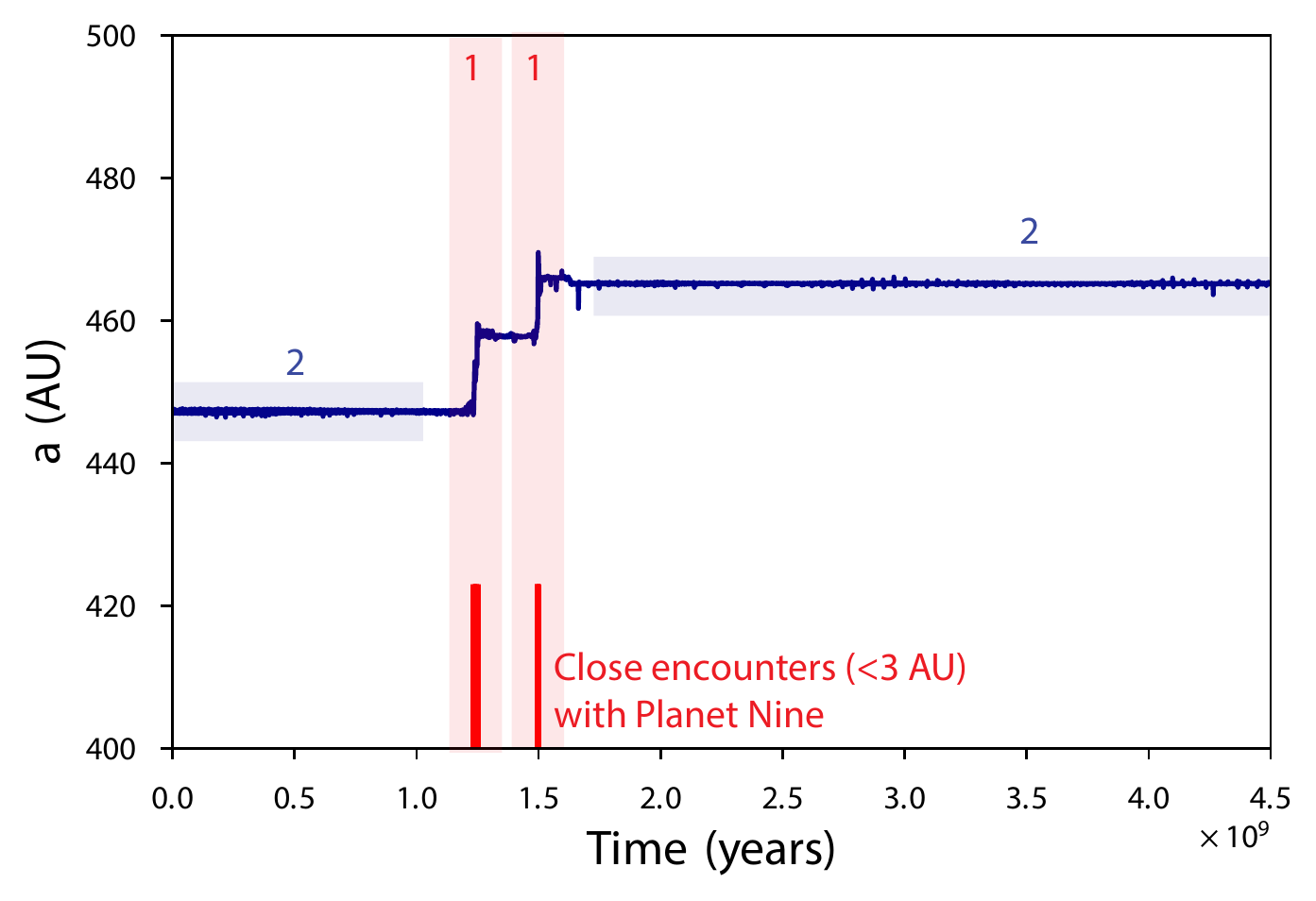}
\caption{A single clone of \thisMainObj\ in the presence of Planet Nine drawn from Set 3 of our simulations. There are two modes of evolution, both shown and labeled in this panel. The first (1) occurs when \thisMainObj\ passes physically close to Planet Nine (a close encounter), and the orbit of the TNO may be slightly jostled. Times when \thisMainObj\ passes within 3 AU of Planet Nine are denoted with red vertical lines. When close encounters occur, the orbit of \thisMainObj\ is altered and appears to migrate for some time before settling into a new equilibrium semi-major axis. These jumps are the same `resonance hopping' discussed in previous work \citep{2017AJ....154...61B}. The second mode of evolution shown here (2) occurs when the semi-major axis remains constant, but the inclination and eccentricity of \thisMainObj\ may still evolve. The work in \citet{2017AJ....154..229B} describes what happens during these (2) regions of constant semi-major axis.  }
\label{fig:example_schematic}
\end{figure}

Many recent papers have considered the existence of a possible ninth planet. 
In this section, we consider how the existence of Planet Nine would alter the orbital behavior and evolution of \thisMainObj. 
In considering possible dynamical interactions between \thisMainObj\ and Planet Nine, there are two main classes of effects that may be relevant:
\begin{itemize}
\item Constant-$a$ evolution (while in or near resonance with another body). Due to \thisMainObj's large semi-major axis, we do not expect Neptune resonances to be relevant. The longest period objects known to be in resonance with Neptune have semi-major axis of $\sim$130 AU \citep{2018arXiv180205805V}.
\thisMainObj's semi-major axis of $\sim$450 AU is likely too large for these processes to be relevant.
However, resonances with Planet Nine may be important.
\item Diffusion and scattering in $a$ due to close encounters with Neptune or Planet Nine. These encounters may be very close ($<$3 AU) and lead to significant changes in the orbit of \thisMainObj, or may be more distant (5-15 AU) and act more as a series of perturbations than an abrupt change. 
\end{itemize}
Both of these modes of evolution can occur over the entire history of the solar system. For example, Fig. \ref{fig:example_schematic} shows one sample numerical realization of the orbital evolution of \thisMainObj, which demonstrates these two evolutionary modes within a single 4.5 Gyr integration. 

\subsection{Evolution with Constant Semi-major Axis}
The existence of Planet Nine can lead to a behavior in which TNOs ``hop'' between resonances \citep{2017AJ....154...61B,2017arXiv171206547H}. This is differentiated from `resonance sticking' \citep{1997Sci...276.1670D,2001Icar..152....4R,2006P&SS...54...87L}, where scattered disk objects are temporarily captured into resonances with Neptune. In the Planet Nine paradigm, TNOs generally spend more time living in resonances than not, with relatively short periods between the attainment of resonances. 

An example of what resonance hopping looks like is given in Fig. \ref{fig:example_schematic}, where the semi-major axis makes sudden transitions between relatively long periods at nearly constant values (note that further examples can be found in Figures 8 and 9 of \citealt{2017AJ....154...61B}). The transitions in the resonance hopping paradigm are generally caused by close encounters with either Neptune or Planet Nine. 

\citet{2017AJ....154..229B} conducted a thorough analytic and numerical exploration of the evolution of TNOs in the case where the TNOs remain at a nearly constant semi-major axis. 
Fig. \ref{fig:example_schematic} demonstrates the typical behavior of \thisMainObj\ in the presence of Planet Nine -- for extended periods of time, it orbits with a roughly constant semi-major axis, until a close encounter (denoted by red vertical bars in the figure) perturbs the semi-major axis into a different value. A new equilibrium is quickly attained, and the object returns to evolution with nearly constant semi-major axis $a$. During the long periods of constant-$a$ orbital motion, the dynamics described in \citet{2017AJ....154..229B} will apply, as described below. 

To study the evolution of \thisMainObj\ under the same conditions starting in the early solar system and integrating to the current day, we conduct another set of simulations (Set 2; see Table \ref{tab:sim_sets_params}). In contrast to the earlier Set 1 integrations, where the giant planets were considered as active bodies, these simulations absorb all four gas giants into the quadrupole moment of the central body. The corresponding contribution of the planets to the value of  $J_{2}$ is given by 
\be
J_{2} = \frac{1}{2} \sum^{4}_{j=1} \frac{m_{j} a_{j}^{2}}{M R_{abs}^{2}}\,,
\label{eq:j2}
\ee
where is $R_{abs}$ is the absorbing radius, within which objects are removed from the simulation, the index $j$ counts through the four gas giants, $m_j$ and $a_j$ denote planetary masses and semi-major axes, and $M$ denotes the mass of the central body. This approximation minimizes perturbations in $a$-space, allowing for an easier study of the orbital evolution at constant-$a$. As was done in \citet{2017AJ....154..229B}, we initialize the inclination of \thisMainObj\ to be drawn from a half-normal distribution with mean 0 degrees and width 5 degrees, which simulates the expected initial conditions in the early solar system. We also include Planet Nine, using the best-fit values of its orbital elements ($a=700$ AU, $e=0.6$, $i=20$, $\omega=150$, $\Omega=90$), which come from \citet{sarah} and \citet{2017AJ....154..229B}. We also run an additional set of simulations (Set 3) with identical parameters, but using the observed inclination of \thisMainObj\, as drawn from the observationally-derived covariance matrix. Simulation Set 2 is intended to study the behavior of an object like \thisMainObj, but starting from early in solar system history, before the inclination of \thisMainObj\ is perturbed to its current-day value. Simulation Set 2 is intended to answer the following question: assuming that \thisMainObj\ started in the same plane as the outer solar system objects that were present in the early solar system, can secular interactions with Planet Nine excite \thisMainObj's inclination to its current day value? For comparison, simulation Set 3 studies the behavior of \thisMainObj\ from the current day forwards (but using the same approximations that are used in Set 2; namely, neglecting perturbations caused by scattering interactions with the giant planets and treating evolution as occurring at constant-$a$). 

\begin{figure}[htbp] 
\centering
\includegraphics[width=3.4in]{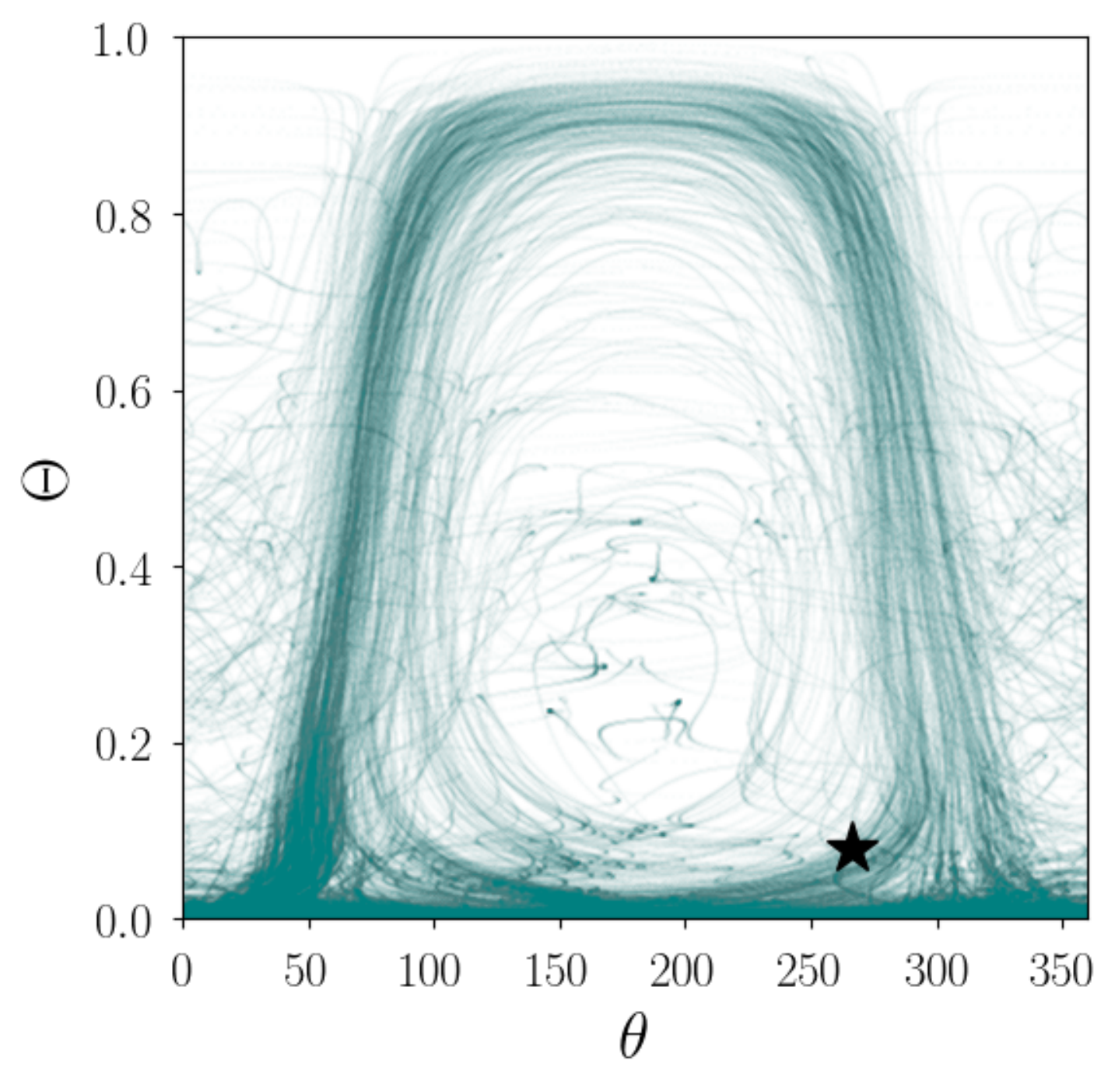}
\caption{Orbital evolution of \thisMainObj\ in action-angle space. This figure shows results from simulations where the initial inclination of \thisMainObj\ was drawn from a half-normal distribution centered at 0 degrees, with a width of 5 degrees (Set 2; see Table \ref{tab:sim_sets_params}). This plot should be compared to the bottom panel of Fig. 11 in \citet{2017AJ....154..229B}. The currently observed action-angle coordinates $\theta$ and $\Theta$ (computed using the simulated version of Planet Nine) is marked by the star symbol.  The current-day orbital elements of \thisMainObj\ are easily reproduced in the scenario with Planet Nine and with \thisMainObj\ starting in the plane with the other solar system objects. }
\label{fig:theta_set3}
\end{figure}

In Fig. \ref{fig:theta_set3}, we plot the action-angle evolution of the results of Set 2, using angle
\be
\theta = \Delta \varpi = 2 \Omega - \varpi - \varpi_{9}
\ee 
and coordinate action 
\be
\Theta = \frac{\sqrt{1-e^{2}}}{2} (1-\cos{i})
\ee
as done in \citet{2017AJ....154..229B}.

The resulting evolution of \thisMainObj\ in this action-angle phase space is plotted in Fig. \ref{fig:theta_set3}. The lines trace the 4.5 Gyr evolution of the realizations of \thisMainObj\ from simulation Set 2. The star symbol marks the present-day location of \thisMainObj\ in this parameter space, using its observed inclination, eccentricity, and (expected) $\Delta \varpi$. It is important to note that the remarkably high observed inclination of \thisMainObj\ is not a guaranteed outcome of these simulations. Nonetheless, the star symbol lies along the teal contours, which describe regions of the phase space to which an initially-coplanar \thisMainObj\ could evolve. These simulations demonstrate that in the case where \thisMainObj\ starts its life close to the plane containing the solar system planets, \thisMainObj\ is able to attain its current day inclination, eccentricity, and orbital orientation through secular interactions with Planet Nine alone. 

As a result, Set 2 of our simulations shows that orbital evolution with constant semi-major axis ($a$) evolution can explain how \thisMainObj\ achieves its observed inclination in the presence of Planet Nine. In other words, the existence of Planet Nine is sufficient to explain the currently observed orbit of \thisMainObj.

\subsection{Orbital Evolution with Planet Nine and Neptune}
\label{sec:Neptune}

The constant-$a$ evolution is relevant for the majority of the lifetimes of the TNOs in the presence of Planet Nine, and the behavior of the TNOs will generally behave as described in the previous section during those times. Close encounters with Planet Nine do occur even in the idealized simulation Set 2, but they are rare and tend to lead to only small hops between nearby resonances with Planet Nine. 
However, as the current perihelion distance of \thisMainObj\ brings it fairly close to the orbit of Neptune during each perihelion passage, the true evolution of \thisMainObj\ will be affected heavily by those Neptune-\thisMainObj\ interactions. In Fig. \ref{fig:example_schematic}, we show a sample orbital evolution of \thisMainObj\ without Neptune. During a close encounter with Planet Nine, \thisMainObj's orbit is rapidly altered, where the average distance of its orbit diffuses until it is trapped into or near a new resonance. The inclusion of Neptune as an active body increases the number of close encounters experienced by \thisMainObj, as it will interact with both Planet Nine and Neptune. This increase in interactions, in turn, allows for the orbit of \thisMainObj\ to become more heavily perturbed over time. 

\begin{figure*}[htbp] 
   \centering
   \includegraphics[width=7in]{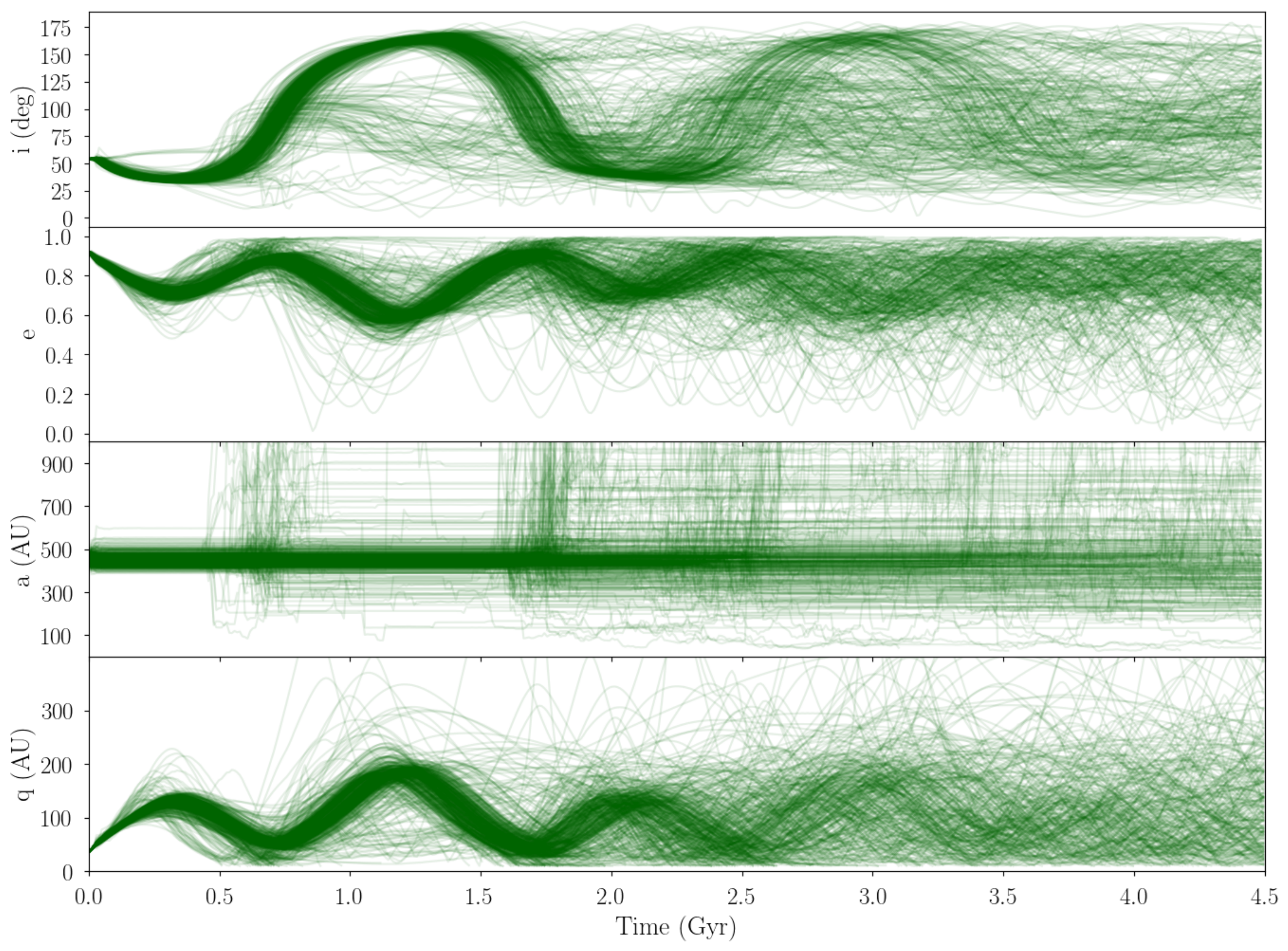}
   \caption{
  The orbital evolution of \thisMainObj\ as computed using N-body simulations for Simulation Set 4, which includes Neptune as an active particle, replaces the other three gas giants with an effective $J_{2}$, and includes Planet Nine. These integrations do describe well the secular dynamics of \emph{surviving} particles. The evolution of semi-major axis also shows the horizontal banding structure in semi-major axis, which is characteristic of resonance hopping.
   }
   \label{fig:set4}
\end{figure*}

To test the effect of these kicks from Neptune, we set up another set of simulations (Set 4; see Table \ref{tab:sim_sets_params}). In this case, we replace Jupiter, Saturn, and Uranus with an effective $J_{2}$ term to represent the potential of those three planets. However, this time we include Neptune as an active body, which allows Neptune scattering events to be resolved. As before, the energy is conserved to one part in 10$^{9}$ and the hybrid symplectic-Bulirsch-Stoer integrator is used. The other parameters of this set of simulations are summarized in in Table \ref{tab:sim_sets_params}, and the results are plotted in Fig. \ref{fig:set4}.
As expected, in this new set of simulations, \thisMainObj\ appears to be significantly less dynamically stable than in the previous sets (which do not include an active Neptune). However, part of this apparent dynamical instability is due to the nonphysical absorbing radius used in the simulations: specifically, we remove particles from the simulation when they reach orbital radii within the absorbing radius. This inner boundary is set to be 9.8 AU in this case, since we are replacing the Sun and inner three giant planets with an oblate central body with a larger radius, to represent the effective quadrupole term of the entire system.

As a result of the complication outlined above, ensembles of simulations that use effective $J_{2}$ terms (like Set 4) to represent time-averaged planetary orbits cannot be used to study the final outcomes of these objects. 
For example, if the orbit of a realization of \thisMainObj\ was to evolve to the point where the clone becomes a Jupiter-family comet, simulation Set 4 would not resolve this end state, and would instead classify the clone as dynamically unstable. On the other hand, this approximation can be used to describe the expected secular evolution for objects that remain a part of the same dynamical population.

\subsection{Orbital Evolution with Planet Nine and the Four Giant Planets}

\begin{figure*}[htbp] 
   \centering
   \includegraphics[width=7in]{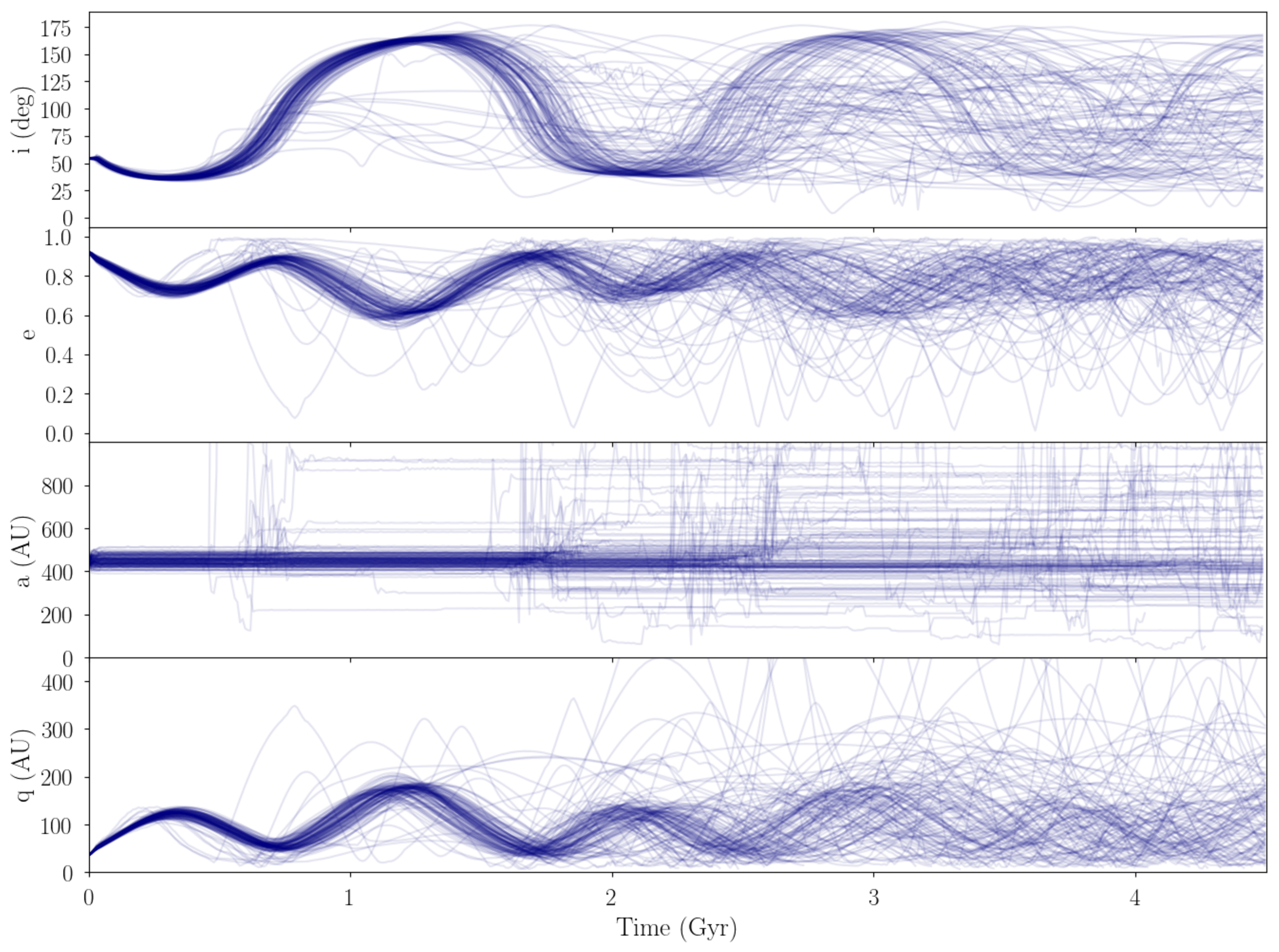}
   \caption{
  The orbital evolution of \thisMainObj\ as computed using N-body simulations for Simulation Set 5, which includes all four gas giants as active particles and also includes Planet Nine. The integrations show the same horizontal banding structure in semi-major axis characteristic of resonance hopping. The evolution computed here is very similar in secular trajectory to that of Set 4 (Fig. \ref{fig:set4}).   }
   \label{fig:set5}
\end{figure*}

Both of the previous sections discussing the orbital evolution of \thisMainObj\ in the presence of Planet Nine replaced some (or all) of the gas giants with an effective $J_{2}$ term. This time-saving integration strategy has been used extensively in the Planet Nine literature \citep{bb16, bb162, sarah, 2017arXiv171206547H}. 
In Section \ref{sec:Neptune}, we showed that the physical presence of Neptune leads to a greater number of transitions (`hops') between Planet Nine (true or near) resonances. Next, our final set of simulations (Set 5) investigates the effect of including all four gas giants as active bodies. The details of Set 5 are given in Table \ref{tab:sim_sets_params}. One important detail about this set of simulations is that since all the gas giants are included as active particles (and terrestrial planets ignored), no planets need be modeled as perturbations on the solar $J_{2}$. As such, the absorbing radius of the central body is set equal to the Solar radius. This aspect of the simulations allows for the resolution of outcomes where \thisMainObj\ settles into a stable orbit with a perihelion distance that passes into the inner solar system; the results of this set of simulations is shown in Fig. \ref{fig:set5}. This figure appears very similar to \ref{fig:set4}, but describes the full motion of \thisMainObj. The striking similarity between the two figures can be used as justification for using the $J_{2}$ approximation when secular evolution is being studied.

\begin{figure*}[htbp] 
\centering
\includegraphics[width=6.8in]{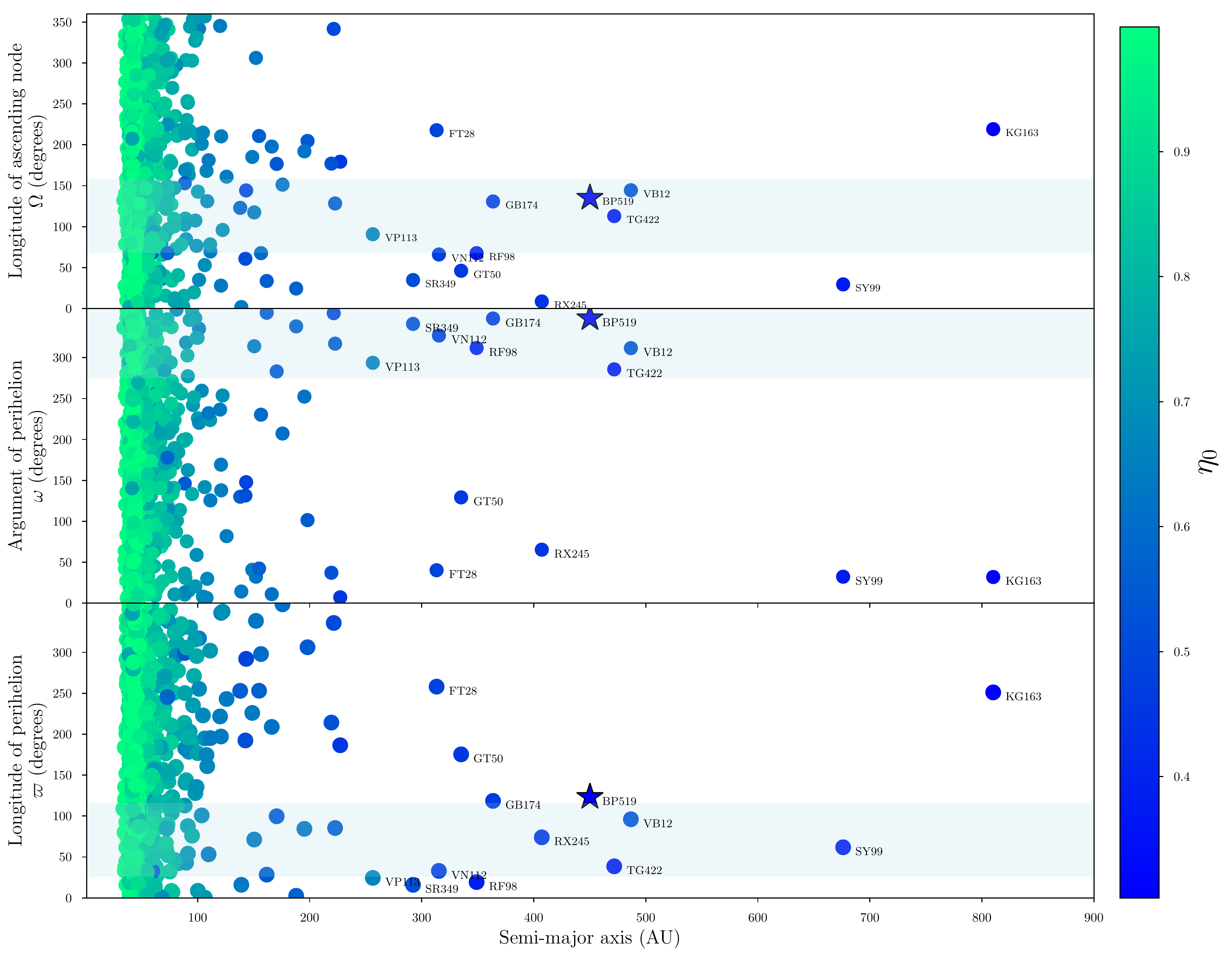}
\caption{A visualization of the two orbital angles $\Omega$ (longitude of ascending node, top panel) and $\omega$ (argument of perihelion, middle panel), along with their sum longitude of perihelion $\varpi=\omega+\Omega$ (top panel). The points are color coded by the specific angular momentum of the orbit $\eta_{0}$ = $\sqrt{1-e^{2}}\cos{i}$. The plot includes all objects with $q>30$ AU and data quality $U < 6$ from the MPC database \citep{1978AJ.....83...64M}, with \thisMainObj\ denoted as a star. Horizontal bars denote the approximate regions of clustering in each angle, as identified in \citet{bb16}.}
\label{fig:all_angles}
\end{figure*}


\begin{figure}[htbp] 
   \centering
   \includegraphics[width=3.4in]{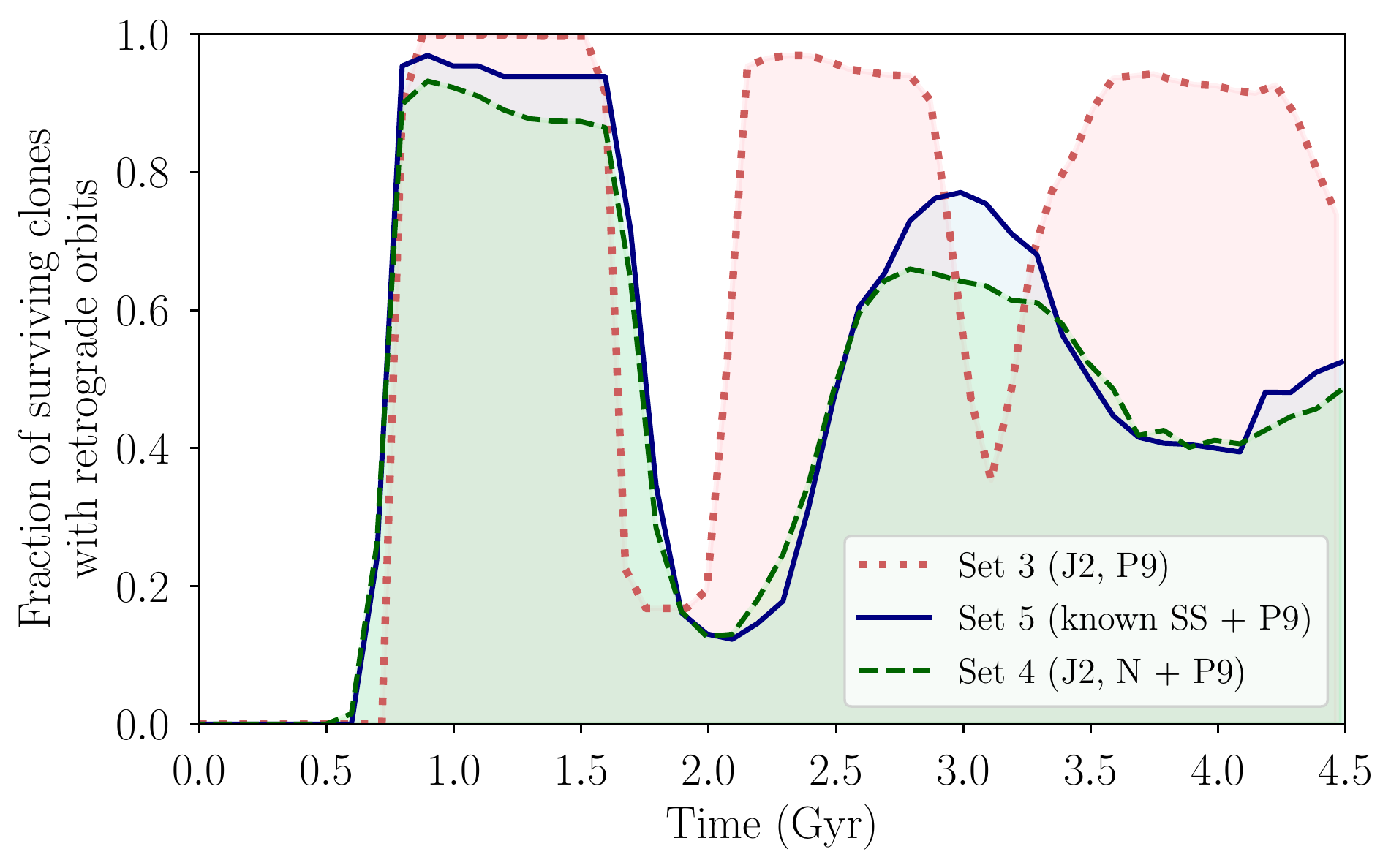}
   \caption{
  A measure of the fraction of \thisMainObj\ clones that attain retrograde orbits as a function of time in the numerical simulations that include Planet Nine. The coherence between all sets of simulations is due to a single realization of Planet Nine's orbital elements being used for all simulations. The good agreement between the simulations that used a $J_{2}$ approximation (Set 4) and those that included all gas giants as active particles (Set 5) suggests that the $J_{2}$ approximation (while keeping Neptune an active particle) is appropriate for studying the orbital evolution of surviving particles, even if it does not work well on its own for studying the dynamical stability. The integrations in Set 1 (which included the known solar system and no Planet Nine) never attain retrograde geometries. 
   }
   \label{fig:retrograde}
\end{figure}

\section{Discussion} 
\label{sec:discussion}

In this work, we present the discovery and dynamical analysis of a new extreme TNO, a population defined as those objects with $a>250$ AU and $q>30$ AU. Because \thisMainObj\ has the largest eccentricity and inclination of any of the extreme TNOs, it allows us to probe the behavior of a new regime in the solar system. Ideally, DES and other surveys will find more of these high-inclination, large semi-major axis objects. Once such a population is found and grows to a sufficient size, it will inform a variety of hypotheses about the structure of the outer solar system and the migration of the giant planets. For now, while the number of such known objects is small, we have performed an in-depth study of the dynamical evolution of \thisMainObj\ in various scenarios with two goals: first, we would like to make whatever insights are possible with a single object to improve our understanding of the outer solar system; second, we would like to determine which hypotheses and analyses will be most fruitful for future study once more of these objects are found. 

Our analysis of the orbital evolution of \thisMainObj\ using forward and backwards integrations has revealed that it is difficult to reproduce \thisMainObj's high current-day inclination in the known solar system without considering some other mechanism. 
In Simulation Set 1, which studied the evolution of this object in the known solar system, 0 out of the 260 simulated clones of \thisMainObj\ attained inclinations less than 48 degrees or greater than 60 degrees, when integrations were initialized with \thisMainObj's measured inclination of $\sim$54 degrees. This strong confinement in inclination space that is evident in the numerical simulations requires us to consider other mechanisms to excite the inclination of this object.
Some potential explanations (discussed in Section \ref{sec:highly_inclined}) include a stellar fly-by, a remnant excitation from the early migration of the giant planets, a particularly serendipitous outcome not captured by our 260 N-body simulations, or an inclination instability caused by the self-gravity of a massive scattered disk.

One additional explanation to those listed above is the existence of a ninth planet in our solar system, as proposed by \citet{st14} and \citet{bb16}. As shown in Fig. \ref{fig:theta_set3}, in the presence of Planet Nine \thisMainObj\ can start out with a relatively low inclination and easily attain its current-day inclination.
Additionally, as shown in Fig. \ref{fig:all_angles}, \thisMainObj's orbital angles $\omega$, $\Omega$, and $\varpi$ appear to be consistent with the clustering first noted in \citet{st14}.
This clustering in physical space has been proposed to be caused by the $\sim$10 Earth-mass Planet Nine at 700 AU \citep{bb16} and is the line of evidence most commonly used to support the existence of Planet Nine.
Although \thisMainObj\ does appear to fit into this paradigm, the physicality of the clustering remains a contentious piece of evidence for Planet Nine \citep{2017AJ....154...50S, 2017AJ....154...65B}.
In Fig. \ref{fig:bias2}, we showed the bias in DES detections of objects with varying orbital angles $\omega, \Omega$, but the same ($a,e,i$) as \thisMainObj\ has.
For \thisMainObj, at least, the biases are sufficiently mild that it seems that \thisMainObj\ can be used as evidence towards the existence of the clustering.
However, the observational biases we determine for \thisMainObj\ do not tell us anything about other objects that may be found by DES or other surveys: without fully accounting for the observational biases for each individual survey that has discovered these ETNOs, it cannot fully be determined how much of the clustering is physical and how much is due to observational bias. 
Past surveys have been able to quantify this: the Deep Ecliptic Survey had well-documented pointings, and as a result was able to construct a model of its detection biases \citep{2014AJ....148...55A}. Similarly, the Outer Solar Systems Origin Survey has quantified its own biases \citep{2018arXiv180200460L}. 
Future work (Hamilton et al., in prep) will do the same for the Dark Energy Survey, and enable a better understanding of whether the clustering suggesting Planet Nine's existence is real or a sampling bias. 

However, \thisMainObj\ does provide additional diagnostics unrelated to angular clustering which inform the Planet Nine debate. 
\citet{bb16} predicted that high inclination KBOs would serve as an important constraint on Planet Nine's properties. Subsequently, dynamical analysis presented in \citet{2017AJ....154..229B} suggests that the population of highly inclined centaurs can be explained by the presence of Planet Nine.
\citet{2017AJ....153...63S} predicted that if there is a ninth planet in the solar system, there should also be a reservoir of high-$i$ TNOs that exhibit clustering of their orbits with the existing population.
Finally, \citet{2017AJ....154..229B} provided a model of the secular evolution expected for high-$i$, high-$a$ objects, but was only able to test it on objects with $q<30$ AU. \thisMainObj\ is the first known high-$a$ ($a>250$ AU), high-$i$ ($i>40$ degrees), high-$q$ ($q >30$ AU) object, a class of objects whose existence is predicted by \citet{2017AJ....154..229B}. 
\thisMainObj\ is the first discovered high-$i$ object, and it fits into the Planet Nine paradigm as predicted by this previous work.

In Fig. \ref{fig:retrograde}, we show the fraction of surviving objects that have retrograde orbits for three of the different simulations sets used in this work.
A sizable fraction of \thisMainObj's potential future orbits attain retrograde orientations, an outcome predicted in \citet{bbincl} and \citet{2017AJ....154..229B}. A subset of these also evolve to lower semi-major axes, potentially resulting in \thisMainObj\ eventually becoming a retrograde centaur; however, our simulations show that it is more likely that \thisMainObj\ retains a large semi-major axis and retrograde configuration than that it migrates inwards and becomes a centaur.  In the presence of Planet Nine, TNOs with orbits as extreme as \thisMainObj\ would appear to cycle though populations, changing their orbital inclinations and perihelion distances rather than living at roughly constant perihelion distances (as they would in the known solar system without Planet Nine; see Fig. \ref{fig:backforth}). Finally, the presence of Planet Nine in the solar system naturally produces objects with orbits like that of \thisMainObj, a feature which cannot be reproduced in the solar system without Planet Nine without invoking some other mechanism (such as interaction with a passing star, or a 1-10 Earth mass scattered disk that can cause an inclination instability, \citealt{2016MNRAS.457L..89M,2018arXiv180503651M}).

Although \thisMainObj\ appears to fit well into the Planet Nine paradigm and aid in a better differentiation between these two potential scenarios -- a solar system with or without Planet Nine -- more objects of this type need to be found. 
Future work using the Dark Energy Survey will both identify additional high-semi-major axis, high-inclination objects which will help us better understand the high-inclination structure of the outermost regions of the solar system, and make a more definitive statement on the existence of Planet Nine. 

\section{Conclusion}
\label{sec:conclude}

This paper reports the detection and initial dynamical analysis of the extreme Trans-Neptunian Object \thisMainObj. This object was discovered as part of the Dark Energy Survey and adds to the growing inventory of unusual bodies in the outer solar system. Our main results can be summarized as follows: 


[1] The estimated orbital elements for this new (minor) member of the solar system include semi-major axis $a\approx$ \thisMainObjSMA, eccentricity $e\approx$ \thisMainObjECC, and inclination $i\approx$ \thisMainObjINC. With these orbital properties, \thisMainObj\ resides well outside the classical Kuiper Belt. On the other hand, the perihelion distance is only $q\sim36$ AU, close enough to be influenced by Neptune. 

[2] The newly discovered body \thisMainObj\ is the most extreme TNO found to date. This claim can be quantified using the reduced Kozai action $\eta_0$ (see Equation \ref{eq:eta}), which is equivalent to the $z$-component of the specific orbital angular momentum. Among all known solar system objects, \thisMainObj\ has the most extreme value of this parameter, as shown in Fig. \ref{fig:eta_level_curves}. 

[3] \thisMainObj\ provides support for the Planet Nine hypothesis. If the object is formed in the plane of the solar system, as expected, then there is a low probability that its orbit can attain the observed high inclination through dynamical processes involving only the known planets. In contrast, the observed 
orbital elements of \thisMainObj\ are readily produced through dynamical interactions if the solar system also contains Planet Nine (see Fig. \ref{fig:theta_set3}). 

\acknowledgements

This material is based upon work supported by the National Aeronautics and Space
Administration under Grant No. NNX17AF21G issued through the SSO Planetary Astronomy Program, and by NSF grant AST-1515015.
We would like to thank Andrew Vanderburg, Ellen Price, Linn Eriksson, Melaine Saillenfest and Mike Brown for many useful conversations. We would like to thank Michele Bannister for useful discussions and methods advice. We would like to thank Konstantin Batygin for his careful review of the manuscript and suggestions that greatly improved this work. 
We thank Marty Kandes and Mats Rynge for help running simulations on Open Science Grid's high-throughput computing resources (operated through XSEDE). J.C.B, S.J.H, and L.M. are supported by the NSF Graduate Research Fellowship Grant No. DGE 1256260. 
This work used the Extreme Science and Engineering Discovery Environment (XSEDE), which is supported by National Science Foundation grant number ACI-1053575. This research was done using resources provided by the Open Science Grid, which is supported by the National Science Foundation and the U.S. Department of Energy's Office of Science, through allocations number TG-AST150033 and TG-AST170008.
This research has made use of data and services provided by the International Astronomical Union's Minor Planet Center.

Funding for the DES Projects has been provided by the U.S. Department of Energy, the U.S. National Science Foundation, the Ministry of Science and Education of Spain, 
the Science and Technology Facilities Council of the United Kingdom, the Higher Education Funding Council for England, the National Center for Supercomputing 
Applications at the University of Illinois at Urbana-Champaign, the Kavli Institute of Cosmological Physics at the University of Chicago, 
the Center for Cosmology and Astro-Particle Physics at the Ohio State University,
the Mitchell Institute for Fundamental Physics and Astronomy at Texas A\&M University, Financiadora de Estudos e Projetos, 
Funda{\c c}{\~a}o Carlos Chagas Filho de Amparo {\`a} Pesquisa do Estado do Rio de Janeiro, Conselho Nacional de Desenvolvimento Cient{\'i}fico e Tecnol{\'o}gico and 
the Minist{\'e}rio da Ci{\^e}ncia, Tecnologia e Inova{\c c}{\~a}o, the Deutsche Forschungsgemeinschaft and the Collaborating Institutions in the Dark Energy Survey. 

The Collaborating Institutions are Argonne National Laboratory, the University of California at Santa Cruz, the University of Cambridge, Centro de Investigaciones Energ{\'e}ticas, 
Medioambientales y Tecnol{\'o}gicas-Madrid, the University of Chicago, University College London, the DES-Brazil Consortium, the University of Edinburgh, 
the Eidgen{\"o}ssische Technische Hochschule (ETH) Z{\"u}rich, 
Fermi National Accelerator Laboratory, the University of Illinois at Urbana-Champaign, the Institut de Ci{\`e}ncies de l'Espai (IEEC/CSIC), 
the Institut de F{\'i}sica d'Altes Energies, Lawrence Berkeley National Laboratory, the Ludwig-Maximilians Universit{\"a}t M{\"u}nchen and the associated Excellence Cluster Universe, 
the University of Michigan, the National Optical Astronomy Observatory, the University of Nottingham, The Ohio State University, the University of Pennsylvania, the University of Portsmouth, 
SLAC National Accelerator Laboratory, Stanford University, the University of Sussex, Texas A\&M University, and the OzDES Membership Consortium.

Based in part on observations at Cerro Tololo Inter-American Observatory, National Optical Astronomy Observatory, which is operated by the Association of 
Universities for Research in Astronomy (AURA) under a cooperative agreement with the National Science Foundation.

The DES data management system is supported by the National Science Foundation under Grant Numbers AST-1138766 and AST-1536171.
The DES participants from Spanish institutions are partially supported by MINECO under grants AYA2015-71825, ESP2015-66861, FPA2015-68048, SEV-2016-0588, SEV-2016-0597, and MDM-2015-0509, 
some of which include ERDF funds from the European Union. IFAE is partially funded by the CERCA program of the Generalitat de Catalunya.
Research leading to these results has received funding from the European Research
Council under the European Union's Seventh Framework Program (FP7/2007-2013) including ERC grant agreements 240672, 291329, and 306478.
We  acknowledge support from the Australian Research Council Centre of Excellence for All-sky Astrophysics (CAASTRO), through project number CE110001020, and the Brazilian Instituto Nacional de Ci\^encia
e Tecnologia (INCT) e-Universe (CNPq grant 465376/2014-2).

This manuscript has been authored by Fermi Research Alliance, LLC under Contract No. DE-AC02-07CH11359 with the U.S. Department of Energy, Office of Science, Office of High Energy Physics. The United States Government retains and the publisher, by accepting the article for publication, acknowledges that the United States Government retains a non-exclusive, paid-up, irrevocable, world-wide license to publish or reproduce the published form of this manuscript, or allow others to do so, for United States Government purposes.

\software{WCSfit (Bernstein et al. 2017), mp\_ephem  (https://github.com/OSSOS/liborbfit), pandas \citep{ mckinney-proc-scipy-2010}, IPython \citep{PER-GRA:2007}, matplotlib \citep{Hunter:2007}, scipy \citep{scipy}, numpy \citep{oliphant-2006-guide}, Jupyter \citep{Kluyver:2016aa}}

\bibliographystyle{apj}

\clearpage
\begin{appendices}

\section{Appendix: Extreme TNO data table}  
\label{app2}

\begin{deluxetable}{lcccccc}[h]
\tablecaption{Orbital Elements of Extreme TNOs}
\tablewidth{0pt}
\tablehead{
  \colhead{Object } &
    \colhead{a (AU) } &
      \colhead{e } &
        \colhead{i } &
          \colhead{$\omega$ } & 
          \colhead{$\Omega$ } &
          \colhead{H (abs. mag) }   \cr
}
\startdata
2003 VB$_{12}$	&	507	$\pm$	10	&	0.8496	$\pm$	0.003	&	11.9	$\pm$	0.1	&	311.3	$\pm$	0.1	&	144.4	$\pm$	0.1	&	1.5	\\
2007 TG$_{422}$	&	503	$\pm$	0.35	&	0.93	$\pm$	0.001	&	18.6	$\pm$	0.1	&	285.7	$\pm$	0.1	&	112.9	$\pm$	0.1	&	6.2	\\
2010 GB$_{174}$	&	351	$\pm$	9	&	0.862	$\pm$	0.004	&	21.6	$\pm$	0.1	&	347.2	$\pm$	0.1	&	130.7	$\pm$	0.1	&	6.6	\\
2012 VP$_{113}$	&	266		$^{+26}_{-17}$	&	0.69	$\pm$	0.03	&	24.1	$\pm$	0.1	&	292.7	$\pm$	0.1	&	90.8	$\pm$	0.1	&	4	\\
2013 FT$_{28}$	&	295	$\pm$	7	&	0.853	$\pm$	0.004	&	17.4	$\pm$	0.1	&	40.7	$\pm$	0.1	&	217.7	$\pm$	0.1	&	6.7	\\
2013 RF$_{98}$	&	363	$\pm$	5	&	0.9	$\pm$	0.001	&	29.6	$\pm$	0.1	&	311.8	$\pm$	0.1	&	67.6	$\pm$	0.1	&	8.7	\\
2013 SY$_{99}$	&	735	$\pm$	15	&	0.932	$\pm$	0.007	&	4.2	$\pm$	0.1	&	32.2	$\pm$	0.1	&	29.5	$\pm$	0.1	&	6.8	\\
2014 SR$_{349}$	&	299	$\pm$	12	&	0.841	$\pm$	0.007	&	18	$\pm$	0.1	&	341.2	$\pm$	0.1	&	34.9	$\pm$	0.1	&	6.6	\\
2015 GT$_{50}$	&	312	$\pm$	2	&	0.877	$\pm$	0.001	&	8.8	$\pm$	0.1	&	129	$\pm$	0.1	&	46.1	$\pm$	0.1	&	8.3	\\
2015 KG$_{163}$	&	680	$\pm$	2	&	0.94	$\pm$	0.001	&	14	$\pm$	0.1	&	32.1	$\pm$	0.1	&	219.1	$\pm$	0.1	&	8.1	\\
2015 RX$_{245}$	&	430	$\pm$	20	&	0.894	$\pm$	0.001	&	12.1	$\pm$	0.1	&	65.2	$\pm$	0.1	&	8.6	$\pm$	0.1	&	6.1	\\
2004 VN$_{112}$	&	316	$\pm$	1	&	0.8505	$\pm$	0.0005	&	25.6	$\pm$	0.1	&	327.1	$\pm$	0.1	&	66	$\pm$	0.1	&	6.5	\\
2014 FE$_{72}$	&	1655 $\pm$ 336	&	0.98 	$\pm$	0.02	&	20.64 	$\pm$	0.1	&	133.89	$\pm$	0.04	&	336.84	$\pm$	0.1	&	6.1	\\
\enddata
\tablecomments{Barycentric osculating elements for the currently known set of TNOs with $a>250$ AU and $q>30$ AU. Excluding our new object 2015 BP519, solutions were drawn from \citet{2017AJ....154...50S} and \citet{2017AJ....153..262B} for all objects except 2013 RF98, 2007 TG422, and 2014 FE72. 
The barycentric orbital solutions for these three objects were fit using the OSSOS \citep{2016AJ....152...70B} implementation\footnote{Available at \url{https://github.com/OSSOS/liborbfit}, and from the Python Package Index via \texttt{pip install mp\_ephem}} of the \citet{2000AJ....120.3323B} orbit fitter, using the observations of each object available\footnote{\url{https://www.minorplanetcenter.net/db_search/} as of 2/1/2018} at the Minor Planet Center.  } 
\label{tab:etno_table}
\end{deluxetable}

\end{appendices}
\clearpage 
\end{document}